\documentstyle[12pt,amsfonts]{article}

\global\arraycolsep=1pt
\begin{document}
\baselineskip 18pt
\begin{titlepage}
%
%
%

\hfill KUL-TF-97/30

\hfill IFUM 592/FT

\hfill {\tt hep-th/9710237}

\hfill  October, 1997

\begin{center}
\hfill
\vskip .4in
{\large\bf On the Evaluation of Compton Scattering Amplitudes in String Theory}

\end{center}
\vskip .4in
\begin{center}
{\large Andrea Pasquinucci~$^{1}$ and Michela Petrini~$^2$}

\vskip .2in

{\sl $^1$ Instituut voor theoretische fysica, K.U.\ Leuven

Celestijnenlaan 200D, B-3001 Leuven, Belgium}

\vskip .14in

{\sl $^2$ Dipartimento di Fisica, Universit\`a di Milano

and INFN, Sezione di Milano

via Celoria 16, I-20133 Milano, Italy}

\end{center}

\vskip1in

\begin{quotation}
\noindent {\bf Abstract}:
We consider the Compton amplitude for the scattering of a photon
and a (massless) ``electron/positron'' at one loop (i.e. genus one)
in a four-dimensional fermionic heterotic string model. Starting
from the bosonization of the world-sheet fermions needed to explicitly
construct the spin-fields representing the space-time fermions,
we present all the steps of the computation which leads to
the explicit form of the amplitude as an integral of modular
forms over the moduli space.

\end{quotation}
\vspace{4pt}
\end{titlepage}
\noindent
\vfill
\eject
\newcommand{\be}{\begin{equation}}
\newcommand{\ee}{\end{equation}}
\newcommand{\ben}{\begin{equation*}}
\newcommand{\een}{\end{equation*}}
\newcommand{\ba}{\begin{eqnarray}}
\newcommand{\ea}{\end{eqnarray}}
\newcommand{\ban}{\begin{eqnarray*}}
\newcommand{\ean}{\end{eqnarray*}}
\newcommand{\brr}{\begin{array}}
\newcommand{\err}{\end{array}}
\newcommand{\bc}{\begin{center}}
\newcommand{\ec}{\end{center}}
\newcommand{\sss}{\scriptscriptstyle}
\newcommand{\bea}{\begin{eqnarray}}
\newcommand{\eea}{\end{eqnarray}}
\newcommand{\bean}{\begin{eqnarray*}}
\newcommand{\eean}{\end{eqnarray*}}
\renewcommand{\theequation}{\thesection.\arabic{equation}}
\def\bbf#1{\mbox{\boldmath $#1$}}
\def\ds#1{\ooalign{$\hfil/\hfil$\crcr$#1$}}
\def\eqmodone{\stackrel{\rm MOD1}{=}}
\def\modone#1{[\hskip-1.2pt[#1]\hskip-1.2pt]}
\def\eqope{\stackrel{\rm OPE}{=}}
\def\ket#1{\vert#1\rangle}
\def\bra#1{\langle#1\vert}
\def\vev#1{\langle#1\rangle}
\def\wew#1{\langle\langle\, #1\, \rangle\rangle}
\def\bivev#1#2{\langle#1\vert#2\rangle}
\def\Teta#1#2{\Theta\left[{}^{#1}_{#2}\right]}
\def\Tetab#1#2{\overline\Theta\left[{}^{#1}_{#2}\right]}
\def\Ical#1#2{{\cal I}\left[{}^{#1}_{#2}\right]}
\def\GGp#1#2{{G^+}\left[{}^{#1}_{#2}\right]}
\def\GGm#1#2{{G^-}\left[{}^{#1}_{#2}\right]}
\def\Teta#1#2{\Theta\left[{}^{#1}_{#2}\right]}
\def\Tetab#1#2{\overline\Theta\left[{}^{#1}_{#2}\right]}
\def\Ical#1#2{{\cal I}\left[{}^{#1}_{#2}\right]}
\def\GGp#1#2{{G^+}\left[{}^{#1}_{#2}\right]}
\def\GGm#1#2{{G^-}\left[{}^{#1}_{#2}\right]}
%
%
\section{Introduction and Summary}
The computation of scattering amplitudes is one of the most powerful
tools we have to study the general features of first-quantized (perturbative)
string theories. These kinds of computations are indeed necessary for a deeper
understanding of the analiticity properties of string amplitudes,
their divergences and associated renormalizations
\cite{DP2}---\cite{Bere}.\par
Even if not of direct phenomenological interest, string amplitudes can
also be useful for our understanding of field theory.
In the low energy limit of string theory ($\alpha'\rightarrow 0$),
the gravitational and non-local effects
are negligible and we get an ordinary field theory.
It is then possible to use the field-theory limit of string scattering
amplitudes to reproduce known results of field theory in an alternative way,
which can lead to the discovery of new features of
field theory itself  (see for instance \cite{Kap,BK,DiV}).\par
Up to now these results (such as new Feynman-like rules for pure Yang-Mills
at one-loop) have been derived using amplitude involving space-time bosons
as external states.
Very few one-loop amplitudes having space-time fermions as external states
have appeared in the literature (see for example ref.\ \cite{At}), mostly
because of some technical issues appearing in the explicit
evaluations of these amplitudes, as discussed for example
in refs.\ \cite{PR0,PR1}. On the other hand, computations of this type
could be very useful if one wishes to get similar results for QCD. \par
In this paper we present one of the simplest four-point
one-loop scattering
amplitude involving external space-time fermions,
that is the Compton scattering
of an ``electron/positron''  and a photon. Here we call ``electron''
(or ``positron'') a massless space-time fermion charged under a $U(1)$
component of the total gauge group.  One can easily extend
the results we present to the case of the scattering of
a massless ``quark'' on a gluon,
since this requires only some simple modifications
of the left-moving part of our equations.\par
We choose to work in a specific four-dimensional heterotic string
model, which has the properties that its
space-time spectrum depends on a set of parameters  and,
as described in ref.\ \cite{PR1}, only for
some values of these parameters is supersymmetric.
Of course the details of the model chosen
for the computation affect the particular features of the scattering
amplitude. However we expect that the general properties of
string scattering amplitudes are independent of the specific string
model, in particular when one takes the field-theory limit.\par
The paper is organized as follows.
In the section 2, we review the KLT-formalism for constructing
four-dimensional heterotic string models with free world-sheet
fermions and we describe the particular model we choose to work with.
In section 3, we list the tools necessary for the
computation of the one-loop amplitude.
By bosonizing the world-sheet fermions,
we introduce the spin-fields with which we build the
vertex operators for the external fermionic states.
We then define the gamma matrices and the charge conjugation matrix.
In section 4 we illustrate the steps of the computation leading to
the ``off-shell''
four-point scattering amplitude of two photons and two massless
chiral fermions. We discuss the role of the PCO operators, the
evaluation of the (world-sheet)
correlators, the appearance of the identities in theta-functions necessary to
have a Lorentz covariant result, and the use of the GSO projection
conditions.
In section 5 we use the Dirac equation and the other on-shell conditions
to obtain the final on-shell
amplitude. Then we discuss the general properties of such amplitude
and we show its independence on the point of insertion of the PCO.
Finally we briefly  discuss the relation of our string scattering
amplitude with the analogous result in field theory.

\section{4d Free Fermion Heterotic String Models}
In this section we briefly review the main lines of the construction
of four-dimensional heterotic string models built with free world-sheet
fermions following the conventions of Kawai-Lewellen-Tye \cite{KLT}
and we describe the particular model we have chosen for the computation
of the one-loop Compton scattering amplitude.
Our notations differ somewhat from those of
ref.\ \cite{KLT}. In particular  we choose to work in the Lorentz-covariant
formulation, rather than in the light-cone gauge. Moreover we perform
all computations directly in  Minkowski  space-time, with metric
$\eta_{\mu\nu}=(-1,1,1,1)$.

\subsection{The KLT Formalism}
The four-dimensional heterotic string models we consider~\cite{KLT}
(see also~\cite{Anto,Bluhm})
are constructed by fermionizing all the two-dimensional
degrees of freedom other than those associated with the
four-dimensional space-time coordinates and by treating them as
free world-sheet fermions.
In the Lorentz-covariant formulation, these models are built with
the four space-time coordinate fields $X^\mu(z, \bar{z})$,
twenty-two left-moving complex fermions
$\bar{\psi}_{(\bar{l})}(\bar{z})$ (with
$\bar{l} = \bar{1}, \ldots , \overline{22}$),
eleven right-moving complex fermions
$\psi_{(l)}(z)$ (with $l= 23, \ldots,33$), right-moving superghosts
$\beta, \gamma$,
left- and right-moving reparametrization ghosts
$\bar{b}, \bar{c}$ and $b,c$.\par

The $N=1$  world-sheet supersymmetry  of the right movers \cite{Anto}
is generated by the supercurrent
\be
T_F = T_F^{[X,\psi]} - c \partial \beta - \frac{3}{2} (\partial c)
\beta + \frac{1}{2}  \gamma b \ , \label{supcurr}
\ee
where the orbital part is given by
\be
T^{[X,\psi]}_F = -\frac{i}{2}\partial X \cdot \psi -\frac{i}{2}
\sum_{m=1}^2 (\psi^m_{(23)}\psi^m_{(24)}\psi^m_{(25)} +
\psi^m_{(26)}
\psi^m_{(27)}\psi^m_{(28)} +
\psi^m_{(29)}\psi^m_{(30)}\psi^m_{(31)}).
\label{supcur}
\ee
In eq.(\ref{supcur}), $\psi^{\mu}$ are four real Majorana fermions related
to the two complex fermions $\psi_{(32)}$ and $\psi_{(33)}$ by:
\bea
\psi^0 &=& \frac{1}{\sqrt2}(\psi_{33} - \psi^*_{33} ),  \quad
\psi^2 = \frac{1}{i\sqrt2}(\psi_{32}  - \psi^*_{32}) ,
\nonumber\\
\psi^1 &=& \frac{1}{\sqrt2}(\psi_{33} + \psi^*_{33} ),  \quad
\psi^3 = \frac{1}{\sqrt2}(\psi_{32}  +  \psi^*_{32}) .
\label{stf}
\eea
They transform as a space-time vector and are
the world-sheet superpartners of the space-time coordinate fields.
Analogously, the real {\it internal} fermions $\psi_{(l)}^{m} (z)$ are
defined from the nine complex right-moving fermions
$\psi_{(l)} (z)$ ($l=23,\ldots,31$) by
\be
\psi^m_{(l)} = \left\{ \frac1{\sqrt2}(\psi_{(l)} + \psi^*_{(l)} )\
,\ \frac1{i\sqrt2}(\psi_{(l)} - \psi^*_{(l)}) \right\}\ ,\quad m=1,2.
\label{intf}
\ee
They correspond to the compactified dimensions and provide internal symmetry
indices to the states of the string.
Moreover, we introduce another right-moving fermion which is needed to
fermionize the superghosts in the usual way,
$\beta=\partial \xi \psi_{(34)}^{*}$
and $\gamma= \psi_{(34)}\eta$. \par
After fermionization, any KLT model is specified by
the set of possible boundary conditions (spin structures) for the 33
world-sheet fermions and  the superghost.
On the cylinder, parametrized by a complex coordinate $z$, the
boundary conditions for the fermions assume the form:
\bea
\bar{\psi}_{(\bar{l})} (e^{-2\pi i} \bar{z})  &=& e^{-2\pi i ({1
\over 2} -
\bar{\alpha}_l)}
\bar{\psi}_{(\bar{l})} (\bar{z}) \ \ \ \  l = 1,\ldots,22 \nonumber\\
\psi_{(l)} (e^{2\pi i} z) &=& e^{2\pi i ({1 \over 2} - \alpha_l)}
\psi_{(l)} (z) \ \ \ \  l= 23,\ldots,33,   \label{bound}
\eea
where $\bar{\alpha}_{l}$ ($l=1,\ldots,22$) and $\alpha_l$ ($l=23,\ldots,33$)
are real numbers.
The Ramond (R) and Neveu-Schwarz (NS) boundary conditions
correspond to $\alpha_{l} = 0$ and $\alpha_l = \frac12$ respectively.\par

World-sheet supersymmetry, modular invariance of the one and multi-loop
partition function \cite{KLT,KLT2}
constraint the possible choices for the fermion boundary conditions.
As a consequence,  any KLT model is completely specified by
a certain number of basis
vectors ${\bf W}_i$, giving the set of possible boundary
conditions (spin structures)
for the fermions, and by a set of parameters $k_{ij}$ defining
the GSO projections.\par
For example, for the supercurrent (\ref{supcurr}) to have
well-defined boundary conditions, the fermions $\psi_{(32)}$ and
$\psi_{(33)}$ associated with space-time coordinates,
the superghosts,
and the products of triplets of internal fermions  must all carry
the same spin structure:
\bea
\sum_{l=23}^{25} \alpha_l
& \eqmodone & \sum_{l=26}^{28}
\alpha_l \eqmodone \sum_{l=29}^{31} \alpha_l \eqmodone \alpha_{32} \nonumber\\
\alpha_{32}
&\ =\  & \alpha_{33} \ =\  \alpha_{34}.
\label{supcurconst}
\eea
Since fermions associated with space-time coordinates can only
have periodic or anti-periodic boundary conditions, it follows
that all the right-moving fermions are restricted to have only R or
NS boundary conditions.
For the left-movers, boundary conditions other than R or
NS are possible. It is this freedom in the choice of the boundary
conditions that allows these models to have interesting gauge groups.\par
Define $\bbf{\alpha}$ to be a 32-dimensional vector of components
$\bar{\alpha}_{l}$ ($l=1,\ldots,22$) and $\alpha_l$ ($l=23,\ldots,32$).
Each vector  $\bbf{\alpha}$ then specifies a choice of boundary
conditions for the free world-sheet fermions as for eq.\ (\ref{bound}).
All vectors compatible with the constraints just described, can
be expressed as linear combinations of a set of basis
vectors ${\bf W}_i$ as \cite{KLT}
\be
\bbf{\alpha}  =
\sum_{i=0,1,\ldots} m_i {\bf W}_i \equiv m {\bf W} , \label{spins}
\ee
where the integers $m_i$ take
values in $\{0,\dots,M_i-1\}$, $M_i$ being the smallest integer
such that $M_i {\bf W}_i$ ($i$ not summed) is a vector of integer
numbers.
The set of basis vectors always includes the vector
\cite{KLT}
\be
 {\bf W}_0\ =\ \left( (\frac12)^{22} \vert
 (\frac12\frac12\frac12)^3(\frac12) \right) ,
\ee
which describes the NS boundary conditions for all fermions.
(Since the fermions $\psi_{(32)}$ and $\psi_{(33)}$ and the superghosts
have the same spin structure  we do not need to extend ${\bf W}_i$
to 34-dimensional vectors by adding $\alpha_{33} = \alpha_{34}$.)\par

Each distinct choice of boundary conditions (each vector
$\bbf{\alpha}$) defines a {\it sector} in the spectrum of
string states. There are $\prod_i M_i$ such
sectors. All states belonging to a given sector have the same
space-time properties: they
are space-time bosons (fermions) depending on whether the last
right-moving component $\alpha_{32}$ of the boundary vector
(which specifies the boundary conditions for the supercurrent)
takes value $1/2$ ($0$) ${\rm MOD}1$.
We will refer to such a sector as a bosonic
or fermionic sector respectively.

In each sector, the set of all possible string
states
is  constructed by acting on the vacuum with the creation operators.
Generally one considers states in the  superghost vacuum with charge
$q' = -1$ ($q' = -1/2$) for a bosonic (fermionic) sector \cite{FMS}.
Physical states are then selected by imposing
the GSO projections, which ensure the correct space-time
statistics of the states.
In the Lorentz-covariant formulation the GSO projections assume
the form \cite{PR1}
\be
{\bf W}_i \cdot {\bf N}_{\modone{\bbf{\alpha}}} - s_i
(N^{(0)}_{\modone{\alpha_{32}}}
 - N^{(\beta \gamma)}_{\modone{\alpha_{32}}}) \eqmodone
\sum_j k_{ij}m_j + s_i + k_{0i} - {\bf W}_i \cdot
\modone{\bbf{\alpha}}.
\label{GSOp}
\ee
Here the inner-product of two vectors, such as ${\bf W}_i \cdot {\bf N}$,
includes a factor of $(-1)$ for right-moving components.
Also, for any real number $\alpha$ we define
$\modone\alpha\equiv\alpha - \Delta$,
where $0\leq\modone\alpha< 1$ and $\Delta \in {\Bbb Z}$;
$s_i = \alpha_{32}$ is the last entry of  ${\bf W}_i $.
${\bf N}_{\modone {\bbf{\alpha}}}$ is the vector of fermion number
operators in the sector $\bbf{\alpha}$,
$N^{(0)}_{\modone{\alpha_{32}}}$ is the
number operator for the ``longitudinal'' complex fermion
$\psi_{(33)}$ and $N^{(\beta \gamma)}_{\modone{\alpha_{32}}}$ is
the superghost number operator
\be
 N^{(l)}_{\modone{\alpha_l}} = \sum_{q=1}^{\infty} \left[
\psi^{(l)}_{-q-
\modone{\alpha_l}+1}\psi^{(l)*}_{q+\modone{\alpha_l}-1} -
\psi^{(l)*}_{-q+\modone{\alpha_l}} \psi^{(l)}_{q-
\modone{\alpha_l}}
\right] , \label{fermnum}
\ee
and
\be
N^{(\beta \gamma)}_{\modone{\alpha_{32}}} = -\sum_{q=1}^{\infty}
\left[
\beta_{-q+\modone{\alpha_{32}}} \gamma_{q-\modone{\alpha_{32}}} +
\gamma_{-q+1-\modone{\alpha_{32}}} \beta_{q-1+\modone{\alpha_{32}}}
\right] ,
\label{sghostnum}
\ee
where we introduced the mode expansions
\bea
\psi_{(l)} (z) & =& \sum_{q \in {\Bbb Z}}
\psi^{(l)}_{q-\modone{\alpha_l}} z^{-q+\modone{\alpha_l}-1/2}
\nonumber\\
\beta(z) & = & \sum_{q \in {\Bbb Z}} \beta_{q-\modone{\alpha_{32}}}
z^{-q+\modone{\alpha_{32}}-3/2} \nonumber\\
\gamma(z) & = & \sum_{q \in {\Bbb Z}} \gamma_{q-\modone{\alpha_{32}}}
z^{-q+\modone{\alpha_{32}}+1/2}. \label{modeexp}
\eea
Consistency at one loop level constrains the quantities $k_{ij}$ parameterizing
the GSO projections eq.\ (\ref{GSOp}) and the vectors
${\bf W}_i$  to satisfy the following conditions
\bea
& k _{ij}&  +k_{ji} \eqmodone {\bf W}_i \cdot {\bf W}_j  \nonumber\\
& M_j & k_{ij} \eqmodone 0   \nonumber\\
& k_{ii} & + k_{i0} + s_i -\frac12 {\bf W}_i \cdot {\bf W}_i
\eqmodone 0. \label{kijeq}
\eea
More precisely, these constraints  follow from the requirement
of modular invariance of the 1-loop
partition function
\bea
{\cal Z} = & & \sum_{m_i,n_j}  C^{\bbf{\alpha}}_{\bbf{\beta}} \int
{{\rm d}^2\tau\over
({\rm Im}\tau)^2 }  \left( \bar\eta(\bar\tau)\right) ^{-24}
\prod_{l=1}^{22} \bar{\Theta}
\left[{}^{\bar{\alpha}_{l}}_{\bar{\beta}_{l}}\right]
(0\vert\bar{\tau}) \nonumber\\
 &  & \times \left(\eta(\tau)\right)^{-12}
\prod_{l=23}^{32} \Theta \left[{}^{\alpha_{l}}_{\beta_{l}} \right]
(0\vert\tau)  {1 \over {\rm Im} \tau} ,\label{partf}
\eea
where the summation coefficients are given by~\cite{PR1,PR2}
\be
C^{\bbf{\alpha}}_{\bbf{\beta}} = {1\over \prod_i M_i}
\exp\left\{- 2\pi
i\left[ \sum_i (n_i + \delta_{i,0})\left( \sum_j k_{ij} m_j
+s_i -k_{i0}\right) + \sum_i m_i s_i +\frac12 \right] \right\}.
\label{phases}
\ee
At one loop it is necessary to specify the boundary conditions of
the fermions around the two non-contractible loops of the torus.
The spin structure $\left[ {}^{\alpha_{l}}_{\beta_{l}}
\right]$ of the fermion $(l)$ is thus parametrized by the two sets
of integers, $m_i$ and
$n_i$, each taking values in $\{0,\ldots,M_i-1\}$:
\bea
 \bbf{\alpha} & = & \sum_{i=0,1,\ldots} m_i {\bf W}_i \nonumber\\
\bbf{\beta}  & = & \sum_{i=0,1,\ldots} n_i {\bf W}_i .
\label{spinstructures}
\eea
The $m_i$ specify the sector of states being propagated in the
loop. The summation over the $n_j$ in eq.\ (\ref{partf})
enforces the GSO projection on the states in the $m_i$'th sector.
Therefore the sum over all spin structures gives the sum over the
full spectrum of GSO projected states circulating in the loop.
The coefficients $ C^{\bbf{\alpha}}_{\bbf{\beta}}$ of eq.(\ref{phases})
are chosen so that all the states in the GSO-projected
spectrum describing space-time bosons (fermions) contribute to the
partition function with weight $+1$ $(-1)$. \footnotemark
\footnotetext{Our expression (\ref{phases}) for the summation
coefficients is somewhat simpler than that given in
ref.~\cite{KLT}, thanks to
certain phases being absorbed into the definition of the $\Theta$
functions (see Appendix A for conventions).}

\subsection{Our Model}
The model we consider has been proposed in
ref.~\cite{KLT} and has been extensively analyzed in ref.~\cite{PR1}.
It has two main features: the gauge group contains
a $U(1)$ and the spectrum can be $N=1$ space-time
supersymmetric or not depending on the values of the
parameters $k_{ij}$. \par

The model is specified by the following boundary
vectors
\bea
 {\bf W}_0 & =& \left( (\frac12)^{22} \vert
(\frac12\frac12\frac12)^3 (\frac12)\right)\nonumber\\
{\bf W}_1 &=& \left( (\frac12)^{22} \vert
(0\frac12\frac12)^3 (0) \right) \nonumber\\
{\bf W}_ 2 &=& \left( (\frac12)^{14} (0)^8 \vert
(0\frac12\frac12) (\frac12 0 \frac12)^2 (0)\right)
\nonumber\\
{\bf W}_3 &=&\left( (\frac12)^7 (0)^7 (\frac12)^3 (0)^5
\vert
(\frac12 0 \frac12) (0\frac12\frac12) (\frac12
\frac12 0) (0)\right)\nonumber\\
{\bf W}_4 &=& \left( (0)^7 (0)^7 (\frac12)^2 (0)
(0)^5\vert
(0\frac12\frac12) (\frac12\frac120) (\frac12\frac12
0) (0)\right)\label{vectors}.
\eea
Since all components of the vectors ${\bf W}_i$ are $0$ or
$1/2$, the integers $M_i$ and $m_i$ assume the values
\be
 M_i\ =\ 2\quad \mbox{and} \quad
 m_i,  n_j \ =\  {0,1}  \quad \mbox{with} \quad i,j=0,\dots , 4.
\ee
The constraints (\ref{kijeq}) are satisfied by
any set of $k_{ij}$ of the form
\be
\pmatrix{k_{00} & k_{01} & k_{02} & k_{03} & k_{04} \cr
k_{10} & k_{11} & k_{12} & k_{13} & k_{14} \cr
k_{20} & k_{21} & k_{22} & k_{23} & k_{24} \cr
k_{30} & k_{31} & k_{32} & k_{33} & k_{34} \cr
k_{40} & k_{41} & k_{42} & k_{43} & k_{44} \cr}
\eqmodone \pmatrix{k_{00} & k_{01} & k_{02} & k_{03} &
k_{04} \cr
k_{01} & k_{01} & k_{12} & k_{13} & k_{14} \cr
k_{02} & k_{12} + \frac12 & k_{02} & k_{23} & k_{24}
\cr
k_{03} & k_{13} + \frac12 & k_{23} & k_{03}+\frac12 &
k_{34} \cr
k_{04} & k_{14}+ \frac12 & k_{24} & k_{34} + \frac12 &
k_{04}+\frac12
\cr} \ ,
\ee
where all the unspecified  $k_{ij}$ take values
${0,\frac12}$, and we choose as independent $k_{00}$ and
$k_{ij}$ with $i<j$. \par
The set of basis vectors (\ref{vectors}) generates a total
of $32$ sectors in the spectrum
according to eq.(\ref{spins}).  To indicate such sectors we
introduce the shorthand notation
\be
\bbf{\alpha} = \sum_{i=0}^4 m_i {\bf W}_i \equiv {\bf
W}_{\rm
subscript},
\ee
where ``subscript'' is the list of those $i$ for which
$m_i = 1$. The only exception is the sector for
which all the $m_i$ are zero which we will just denote
by $\bbf{\alpha}= {\bf 0}$. \par
{}From the left-moving part of the vectors (\ref{vectors}),
it follows that the world-sheet fermions are grouped
together according to ${\bf W}_4$: For example, the first seven
left-moving complex fermions always have the same spin
structure; from the corresponding $14$ real fermions we may build up
the Ka\v{c}-Moody algebra of $SO(14)$. Therefore the gauge group of
the model is
\be
SO(14)\otimes SO(14) \otimes SO(4) \otimes U(1) \otimes
SO(10),
\label{group}
\ee
where the $U(1)$  is actually realized as an $SO(2)$.
Indeed, as we will show presently, the gauge bosons
in the physical spectrum exactly fill out the adjoint
representation of this group. \par
In ref.\ \cite{PR1} the reader can find the list of the
vacuum energies of all the 32 sectors
in the model and of the excited states up to mass level $\alpha' M^2 = 1$,
as well as the truncation of the spectrum due to the GSO projections.
Here we just review the main features of the spectrum which will
interest us.\par
In each sector, the vacuum state is given by the tensor product of the vacua
corresponding to every single fermion.
Since the vectors ${\bf W}_i$
have all entries $0$ and $1/2$, each  fermion can only have
Ramond or Neveu-Schwartz boundary
conditions. In the first case, the vacuum is the conformal one,
$\vert 0 \rangle$, while in the second it is given by the two-fold degenerate
Ramond vacua $\vert a_{l}\rangle$, where $a_{l} = \pm 1/2$.\par
For a set of several Ramond vacua we introduce the shorthand notation
\newline $\vert a_{24,28,29},\alpha \rangle \equiv
\vert a_{24},a_{28},a_{29},\alpha \rangle$,
where $\alpha \equiv (a_{32},a_{33})$ is a space-time spinor index.\par
The spectrum of excited states is constructed, sector by
sector, by acting on the vacuum with the creation
operators. We restrict ourselves to states in the
(super) ghost vacuum with superghost charge $q = -1$
($q =-1/2$) for bosonic (fermionic) sectors.
Only states satisfying the level-matching condition
$L_0 = \bar{L}_0$ can propagate, and their masses are given by
\be
\frac{\alpha^{\prime}}{4} M^2 = \bar{L}_0 -
\frac{\alpha^{\prime}}{4} p^2 =
L_0 -\frac{\alpha^{\prime}}{4}p^2 .
\ee
In terms of the oscillator operators for the various
excitation modes, the mass of a state in a sector
$\bbf{\alpha}$ is \cite{KLT,PR1}
\bea
\frac{\alpha^{\prime}}{4} M^2  & = & \sum_{l=23}^{33}
\left\{ E_{\modone{\alpha_l}} + \sum_{q=1}^\infty
\left(
( q+\modone{\alpha_l}-1) \,
n^{(l)}_{q+\modone{\alpha_l}-1}
\right.\right. \nonumber\\
& & + \left.\left.  (q - \modone{\alpha_l}) \,
n^{(l)*}_{q-\modone{\alpha_l}}
\right) \right\} + \sum_{q=1}^\infty
q a_{-q} \cdot a_q -1 + E^{(\beta
\gamma)}_{\modone{\alpha_{32}}},
\eea
where
$a_q^{\mu}$ are the (right-moving) modes of
$X^{\mu}(z,\bar{z})$, $n^{(l)}$ and $ n^{(l)*}$ are the
fermion and antifermion mode occupation numbers defined
by
\be
 n^{(l)}_{ q+\modone{\alpha_l}-1}= \psi^{(l)}_{-q-
\modone{\alpha_l}+1}
\psi^{(l)*}_{q+\modone{\alpha_l}-1}, \quad
 n^{(l)*}_ {q-\modone{\alpha_l}}= \psi^{(l)*}_{-
q+\modone{\alpha_l}}
 \psi^{(l)}_{q-\modone{\alpha_l}}.
\ee
$E^{(\beta \gamma)}_{\modone{\alpha_{32}}}$ is the
superghost vacuum energy,
which equals $+1/2$ ($+3/8$) in a bosonic (fermionic)
sector, while the contribution of minus one represents the
reparametrization ghost vacuum energy. Finally
$E_{\modone{\alpha_l}}$ is  the vacuum energy of
the $l$'th complex fermion (relative to the conformal vacuum)
\be
E_{\modone{\alpha_l}}=\frac12\left(\modone{\alpha_l}-
\frac12\right)^2.\label{vacenergy}
\ee
A similar formula holds for the left-movers, without
the superghost vacuum
energy, and with right-movers replaced by left-movers.\par
Let us consider first the bosonic sector ${\bf W}_0$.
Before the GSO projections, it contains the standard charged
tachyon of mass squared $\alpha' M^2 = -2$, and some massless states:
the graviton, dilaton and axion
$$
\bar{a}_{-1}^{\mu} \ket{0}_L \otimes
\psi_{-1/2}^{\nu}\ket{0}_R\ ,
$$
a set of charged vectors
\bean
\bar\psi_{-1/2,(\bar{l})}^{\overline{m}}
\bar\psi^{\overline{n}}_{-1/2,(\bar{k})} \ket{0}_L &\otimes&
\psi_{-1/2}^{\mu}
\ket{0}_R \quad \bar{l}, \bar{k}=1,\dots,22, \bar{l}\neq\bar{k},\\
\bar{a}_{-1}^{\mu} \ket{0}_L &\otimes&
\psi_{-1/2, (j)}^{m}\ket{0}_R  \quad j=23,\dots,31\ ,
\eean
and a set of charged scalars
$$
\bar\psi_{-1/2,(\bar{l})}^{\overline{m}}
\bar{\psi}^{\overline{n}}_{-1/2,(\bar{k})} \ket{0}_L \otimes \psi_{-1/2,
(j)}^m \ket{0}_R \quad \bar{l}, \bar{k}=1,\dots,22, \bar{l}\neq\bar{k},
j=23,\dots,31.\footnotemark
\footnotetext{$m,n=1,2$ as in eq.(\ref{intf})}
$$
The GSO projections eliminate from the physical spectrum the tachyon, all
scalars, as well as the spin-$1$ states where the vector index is carried
by the oscillators $\bar{a}_{-1}^{\mu}$. Moreover the massless
vectors fill out the adjoint representation of the gauge group
$SO(14)\otimes SO(14) \otimes SO(4) \otimes U(1) \otimes SO(10)$,
and therefore are the gauge bosons of the model.\par
In the sector ${\bf W}_1$ we find 8 massless spin-$1/2$ states
$$
\bar\psi_{-1/2,(\bar{l})}^{\overline{m}}
\bar\psi^{\overline{n}}_{-1/2,(\bar{k})} \ket{0}_L \otimes
\ket{a_{23,26,29},\alpha}_R \quad \bar{l}, \bar{k}=1,\dots,22,
\bar{l}\neq\bar{k}
$$
and 8 massless spin-$3/2$
$$
\bar{a}_{-1}^{\mu} \ket{0}_L \otimes
\ket{a_{23,26,29},\alpha}_R,
$$
which represent the gauginos  and gravitinos respectively.
Imposing the  GSO projections, one finds that only a single
gravitino survives, if and only if
\be
k_{02}+ k_{12} \eqmodone k_{04} + k_{14}.
\label{Susyc}
\ee
This is then the condition for the model to be $N=1$ supersymmetric.
The same analysis applies to the gauginos, leading consistently
to the same condition for space-time supersymmetry. \par
If the model is supersymmetric (i.e. equation (\ref{Susyc}) holds), then
given a state in the sector $m{\bf W}$, the
superpartner resides in the sector
${\bf W}_0 + {\bf W}_1 + m{\bf W}$ ~\cite{PR1}.
Notice that ${\bf W}_{\rm SUSY}={\bf W}_0 + {\bf W}_1$ also exchanges
the boundary conditions of
the internal world-sheet fermions $\psi_{(23)}$, $\psi_{(26)}$ and
$\psi_{(29)}$.
The associated degrees of freedom are not family indices for the states
and should be considered instead as enumerative indices for the elements
of the space-time supermultiplets.\par

We  end this section by considering
in some details the GSO projection conditions for the ground states
of the sector ${\bf W}_{134}$.
In this sector, the ground state is given by a set of massless fermions
charged under the $U(1)$ and the first $SO(14)$ components
of the gauge group:
$\ket{\bar{a}_{1,\dots,7,17}}_L\otimes\ket{a_{24,28,29},\alpha}_R$.
These are the fermions which we will use in the Compton amplitude.
On such states the GSO conditions, eq.(\ref{GSOp}),
become (considering only zero-mode excitations)
\bea
& \frac12 &\left[
\sum_{l=1}^{7} \bar{N}^{(\bar{l})}_{0} + \bar{N}^{(\bar{17})}_{0}
- \sum_{l=24,28,29,32,33} N^{(l)}_{0} \right]
\eqmodone \frac12 + k_{00} + k_{01} + k_{03} + k_{04}, \label{GSOone}\\
&\frac12& \left[
\sum_{l=1}^{7} \bar{N}^{(\bar{l})}_{0} + \bar{N}^{(\bar{17})}_{0}
- \sum_{l=24,28} N^{(l)}_{0} \right]
\eqmodone -\frac12 + k_{13} + k_{14}, \label{GSOtwo}\\
&\frac12& \left[
\sum_{l=1}^{7} \bar{N}^{(\bar{l})}_{0}- \sum_{l=24,28,29} N^{(l)}_{0} \right]
\eqmodone \frac12 + k_{02}+ k_{12}+ k_{23} + k_{24}, \label{GSOthree}\\
&\frac12& \left[
\sum_{l=1}^{7} \bar{N}^{(\bar{l})}_{0} + \bar{N}^{(\bar{17})}_{0}
-\sum_{l=28,29} N^{(l)}_{0} \right]
\eqmodone \frac12 + k_{13} + k_{34}, \label{GSOfour}\\
&\frac12& \left[-\sum_{l=24,29} N^{(l)}_{0} \right]
\eqmodone  k_{14} + k_{34}. \label{GSOfive}
\eea
It is convenient to rewrite the GSO conditions for the Ramond zero modes
in term of Pauli matrices.
Consider the generic projection condition
\be
\frac12 \left(\sum_{l\in \bar{I}} \bar{N}^{(l)}_0
- \sum_{l \in I} N^{(l)}_0 \right) \eqmodone r,
\ee
where $r \in \{0,1/2\}$ and the left-hand side involves a total of $m$
number operators. Then, since $N^{(l)}_0 = \frac12 (1+\sigma^{(l)}_3)$
for zero-mode excitations,
this projection condition can be rewritten as
\bea
\bigotimes_{l\in \{\bar{I},I\}} \sigma^{(l)}_3 &=&
\exp\left\{2\pi
i [ r +\frac12 ] \right\} \qquad {\rm for} \qquad
m\ {\rm odd} \nonumber\\
\bigotimes_{l\in \{ \bar{I},I\}} \sigma^{(l)}_3 &=& \exp\left\{2\pi i
r \right\} \qquad\qquad {\rm for} \qquad
m\ {\rm even}.
\eea
Then the projection conditions eqq.(\ref{GSOone}-\ref{GSOfive}) become
\bea
\sigma_3^{(\overline{17})}\otimes \gamma^5 &=&
\exp\left\{2\pi i
\left[k_{00} + k_{01} + k_{02}+ k_{03} + k_{04} + k_{12}
 + k_{23} + k_{24}+ \frac12 \right]\right\} \nonumber\\
\sigma_3^{(24)}\otimes \gamma^5 &=&
\exp\left\{2\pi i
\left[k_{00} + k_{01} + k_{03} + k_{04}
 + k_{13} + k_{34}+ \frac12 \right]\right\} \nonumber\\
\sigma_3^{(29)}\otimes \gamma^5 &=&
\exp\left\{2\pi i
\left[k_{00} + k_{01} + k_{03} + k_{04}
 + k_{13} + k_{14}+ \frac12 \right]\right\} \nonumber\\
\gamma_{SO(14)} \otimes \sigma_3^{(28)} &=&
\exp\left\{2\pi i
\left[k_{02} + k_{12} + k_{14} + k_{23}
 + k_{24} + k_{34}+ \frac12 \right]\right\},
\eea
where we introduced the space-time chirality operator
$\gamma^5 \equiv\sigma_3^{(32)} \otimes \sigma_3^{(33)}$ and
the chirality matrix in the spinor representation of
the gauge group $SO(14)$, $\gamma_{SO(14)} =
\bigotimes_{l=1}^7\sigma_3^{(\bar{l})}$.

\section{Amplitudes, Vertex Operators and Cocycles}
\setcounter{equation}{0}
In this section we introduce the tools necessary for the computation
of the one-loop Compton scattering amplitude in the context of a KLT
four-dimensional model in a Minkowski background following refs.\
\cite{PR1,PR2,PR3}.\par
We define the $T$-matrix element as the connected $S$-matrix
element with certain normalization factors removed
\bea
\lefteqn{\frac{ \langle \lambda_1, \dots , \lambda_{N_{\rm out}} \vert S_c
\vert
\lambda_{N_{\rm out}+1}, \dots , \lambda_{N_{\rm out} + N_{\rm
in}}\rangle}
{\prod_{i=1}^{N_{\rm tot}} \left( \langle \lambda_i \vert
\lambda_i \rangle \right)^{1/2} }  = } \nonumber\\
   & & i (2\pi)^4 \delta^4 (p_1 + \dots + p_{N_{\rm out}}-
p_{N_{\rm out}+1}- \dots -
p_{N_{\rm tot}})
   \prod_{i=1}^{N_{\rm tot}} (2 p^0_i V)^{-1/2} \times \nonumber\\
& & T(\lambda_1, \dots , \lambda_{N_{\rm out}} \vert
\lambda_{N_{\rm out}+1}, \dots , \lambda_{N_{\rm out} + N_{\rm in}} ),
 \label{Smatrix}
\eea
where $N_{\rm tot} = N_{\rm in} + N_{\rm out}$ is the total number
of external states, $p_i $ is the momentum of the $i$'th string state
($p^0_i > 0 $ for all states) and $V$ is the usual
volume-of-the-world factor.\par
Corresponding to each state $\vert \lambda \rangle$ we have a
vertex operator ${\cal V}_{\vert \lambda \rangle} (\bar{z},z)$ and
the 1-loop contribution to the $T$-matrix element is given by the
operator formula
\bea
 T^{1-{\rm loop}}
&(& \lambda_1, \dots , \lambda_{N_{\rm out}} \vert
\lambda_{N_{\rm out}+1}, \dots , \lambda_{N_{\rm out} + N_{\rm
in}} ) \ = \nonumber\\
& & C_{g=1} \int \prod_{I=1}^{N_{\rm tot}}
{\rm d}^2 m^I \ \sum_{m_i,n_j}  C^{\bbf{\alpha}}_{\bbf{\beta}}
\langle \langle \left| \prod_{I=1}^{N_{\rm tot}} (\eta_I \vert b)
\prod_{i=1}^{N_{\rm tot}} c(z_i) \right|^2 \nonumber\\
& & \prod_{A=1}^{N_B+N_{FP}}
\Pi (w_A)\ {\cal V}_{\langle \lambda_1 \vert} (\bar{z}_1,z_1)
\dots
{\cal V}_{\vert \lambda_{N_{\rm tot}} \rangle}
 (\bar{z}_{N_{\rm tot}}, z_{N_{\rm tot}}) \rangle \rangle.
\label{Tmatrix}
\eea
Here $C_{g=1} =1/(2 \pi \alpha')^2$ is a constant
giving the correct normalization of the vacuum amplitude \cite{PR2,Kaj}.
The coefficients  $C^{\bbf{\alpha}}_{\bbf{\beta}}$
of the summation over spin structures are given in eq.\ (\ref{phases}).
$m^I$ is a modular parameter, $\eta_I$ is the corresponding
Beltrami differential~\cite{DPh}, and the overlap $(\eta_I \vert b)$
with the antighost field $b$ is given explicitly in ref.~\cite{Kaj2}.
The integral is over one
fundamental domain of $N$-punctured genus-one moduli space.
By definition the correlator $\langle \langle{\dots}\rangle \rangle$
includes the partition
function (more details on our conventions for the partition
function, spin structure and field operators can be found
in Appendix A, see also ref~\cite{PR1,PR2,PR3}).

In eq.\ (\ref{Tmatrix}), the ghost factors present in the BRST
invariant version of the vertex operator,  ${\cal W}_{\vert \lambda
\rangle} (\bar{z},z) =  \bar{c}(\bar{z})c(z){\cal V}_{\vert
\lambda \rangle} (\bar{z},z)$, have been factored out.
One generically takes all space-time fermionic vertex
operators to have the superghost charge $q=-1/2$  and all the
bosonic ones to have the superghost charge $q=-1$.
In an amplitude involving $N_B$ space-time bosons and $2N_{FP}$
space-time fermions, in order to compensate the superghost vacuum charge,
we insert $N_B + N_{FP}$
Picture Changing Operators (PCO) $\Pi(w_A)$, at
arbitrary points $w_A$ on the Riemann surface ~\cite{FMS, DPh}.
In practical calculations
it is more convenient to insert one PCO at each of the vertex
operators describing the space-time bosons. This leaves $N_{FP}$
PCO's at arbitrary points and the boson vertex operators in their picture
changed version with superghost charge $q=0$.\par

An explicit expressions of the vertex operators and of the PCO
can be given bosonizing all the fermionic
degrees of freedom. We use the
prescription for bosonization of world-sheet fermions
in Minkowski space-time proposed in ref.\ \cite{PR3}.
Bosonization in Minkowski space-time differs from the one in
Euclidean space-time because the world-sheet fermions
$\psi_{(33)}$ and $\psi_{(33)}^*$, which are related to the time
direction in space-time, have different hermiticity properties
compared to all other fermions. All left- and
right-moving complex fermions are bosonized according to
\bea
\psi_{(l)}(z)  &=&  e^{\phi_{(l)}(z)}c_{(l)} \qquad\qquad\qquad
\psi_{(l)}^*(z) = e^{-\phi_{(l)}(z)} c^*_{(l)} \nonumber\\
\bar{\psi}_{(\bar{l})}(\bar{z}) & = &
e^{\bar{\phi}_{(\bar{l})}(\bar{z})}
c_{(\bar{l})} \qquad\qquad\qquad
\bar{\psi}_{(\bar{l})}^*(\bar{z})
= e^{-\bar{\phi}_{(\bar{l})}(\bar{z})}
c^*_{(\bar{l})}, \label{bosonization}
\eea
where the scalar field $\phi_{(l)}$ has operator product expansion
(OPE)
\be
\phi_{(l)}(z) \phi_{(k)}(w) = +  \delta_{l,k} \log (z-w) + \ldots
\ . \label{OPEscalars}
\ee
The factors $c_{(l)}$ are cocycles needed to guarantee the correct
anti-commutation relations between different fermions.
We will return to them in the next subsection.\par
As shown in~\cite{PR3}, because of the Minkowski metric in the OPE,
\be
\psi^\mu(z) \psi^\nu(w) \eqope
\frac{g^{\mu \nu}}{z-w} +\ \cdots,
\ee
the fermion $\psi^0$ (see eq.\ (\ref{stf})) and the associated scalar field
$\phi_{(33)}$ are
hermitian, while all other fermions, with fields $\phi_{(l)}$
($\bar{l}=1,\ldots,22$ and $l=23,\ldots,32$), are anti-hermitian.\par
The superghosts are bosonized in the standard way
\be
\beta\ =\  \partial \xi \, e^{-\phi} (c_{(34)})^{-1} \ ,\qquad\qquad
 \gamma\ =\ e^{\phi} c_{(34)} \  \eta\ , \label{bossupgh}
\ee
where $c_{(34 )}$ is their cocycle factor.
The scalar field $\phi$ in (\ref{bossupgh}) is hermitian and
has the ``wrong'' metric
\be
 \phi(z) \phi(w) = - \log (z-w) + \ldots \ , \label{supghOPE}
\ee
and the PCO is given by
\be
\Pi = 2c\partial \xi + 2 e^\phi c_{(34)}T^{[X,\psi]}_F -
\frac12 \partial (e^{2\phi} (c_{(34)})^2 \eta b)
-\frac12 e^{2\phi}(c_{(34)})^2 (\partial \eta) b.
\label{pco}
\ee
The spin field operator  which creates from the conformal vacuum
the Ramond ground state  associated with the $l$'th fermion can be
written as
\be
S^{(l)}_{a_l}(z) = e^{a_l \phi_{(l)}(z) }
(c_{(l)})^{a_l},\label{bosspinf}
\ee
where  $a_l = \pm \frac12 $ is related to the spin structure
$\alpha_l$ by $ a_l = \frac12 - \alpha_l
\ {\rm MOD} \ 1$. A similar expression holds for the left-movers.\par
If we define the scalar field in (\ref{bossupgh}) as  $\phi\equiv
\phi_{(34)}$, the superghost part of any physical state vertex
operator can be expressed  by eq.(\ref{bosspinf}) with $l=34$ and
\be
a_{34} = -\frac12 - \modone{\alpha_{32}} = \left\{ \begin{array}{ll}
 -1 & \mbox{ in bosonic sector} \\
 -1/2&  \mbox { in fermionic sector.}
\end{array}
\right. \label{supghcharge}
\ee
Then, in any given sector $\bbf{\alpha}$ the vertex operator
describing the
ground state of momentum $p$ has the form
\be
 {\cal V} = {\cal N} \cdot \prod_{\bar{l} =
\bar{1}}^{\overline{22}}
\bar{S}^{(\bar{l})}_{\bar{a}_l}
\prod_{l=23}^{34}  S^{(l)}_{a_l} \cdot  e^{i k\cdot X}
\equiv {\cal N} \cdot S_{\bf A} \cdot e^{i k\cdot X},
\label{vertop}
\ee
where ${\cal N}=\kappa / \pi$ \cite{PR2},
${\bf A} \equiv (A;a_{34}) \equiv
(\bar{a}_1,\ldots,\bar{a}_{22};a_{23},\ldots;a_{34})$ and
we introduced the dimensionless momentum
$k_\mu \equiv \sqrt{{\alpha^\prime \over 2}} \, p_\mu$. \par
\subsection{Cocycles}
As already mentioned, the cocycle factors are necessary to ensure
that different fermions anti-commute when they are written in the
bosonized form.
We write the cocycle operators as follows~\cite{PR1,PR3}
\bea
c_{(\bar{l})} & = & c_{\rm gh}^{(\bar{l})} \cdot
\exp \left\{ i \pi \sum_{j=1}^{l-1} Y_{\bar{l} \bar{j}}
\bar{J}^{(\bar{j})}_0  \right\}   \qquad {\rm for}
\qquad \bar{l} = \bar{1},\dots,\overline{22}  \label{cocycles}\\
c_{(l)} & = & c_{\rm gh}^{(l)} \cdot
\exp \left\{ i \pi \left( \sum_{j=1}^{22} Y_{l \bar{j}}
\bar{J}^{(\bar{j})}_0 + \sum_{j=23}^{l-1} Y_{lj} J^{(j)}_0 \right)  \right\}
\qquad {\rm for}
\qquad l = 23,\dots,34 , \nonumber
\eea
with
\bea
c_{\rm gh}^{(l)}  &\equiv &\exp\{ -i \pi \varepsilon^{(l)} N_{(\eta,\xi)} \}
\exp \{ i \pi \varepsilon^{(l)} (N_{(b,c)} - N_{(\bar{b},\bar{c})}) \}
\nonumber\\
c_{\rm gh}^{(\bar{l})} &\equiv& \exp\{ -i \pi \varepsilon^{(\bar{l})}
N_{(\eta,\xi)} \}
\exp \{ i \pi \varepsilon^{(\bar{l})} (N_{(b,c)} - N_{(\bar{b},\bar{c})})
\}.
\label{cgh}
\eea
Here all the parameters $Y$, as well as the $\varepsilon$,
take values  $+1$ or $-1$.
$N_{(b,c)}$, $N_{(\bar{b},\bar{c})}$ and $N_{(\eta,\xi)}$ are the
number operators of the $(b,c)$, $(\bar{b},\bar{c})$ and $(\eta,\xi)$
systems respectively, while
$\bar{J}^{(\bar{j})}_0$ and $J^{(j)}$ are the number operators
for the world-sheet fermions
\be
J_0^{(l)}  = \oint_0 {dz \over 2\pi i} \partial \phi_{(l)} (z)
\ee
where $\partial \phi_{(l)}  =  - \psi_{(l)}^{*} \psi_{(l)}
= - i \psi_{(l)}^1 \psi_{(l)}^2$.\par
A priori there are many possible different choices of the cocycle operators,
which reflect themselves in the arbitrarity in the signs
$Y$ and $\varepsilon$.
The possible choices are however restricted by a certain number of
consistency conditions ~\cite{PR1,Koste}:
Cocycles have to be chosen in
such a way that the left- and right-moving BRST currents have well-defined
statistics with respect to all vertex operators, otherwise,
a product of BRST invariant vertex operators would not
necessarily be BRST invariant.
Likewise, all
Ka\v{c}-Moody currents must satisfy Bose statistics with respect to all
vertex operators, otherwise, a product of vertex operators ${\cal
V}_i$ transforming in various representations $D_i$ of the gauge group
would not necessarily transform in the tensor representation $\otimes_i
D_i$. It is also necessary that the PCO eq.\ (\ref{pco})
obeys Bose statistics with respect to all vertex operators.
These consistency conditions have been discussed
in the case of Euclidean space-time in ref~\cite{PR1}, where
it is also shown how to construct an explicit
solution to these constraints for our specific model.
The same conditions hold for Minkowski space-time~\cite{PR3},
thus the discussion in ref.\ \cite{PR1} applies also to our
case. \par
In the computation that follows, we will not use any specific choice
of the set of cocycles since, as follows from ref.\ \cite{PR1},
a choice of cocycles which satisfies all consistency conditions
exists and all choices of cocycles that satisfy the consistency
conditions give rise to the same scattering amplitude. In other words,
the consistency conditions guarantee that the scattering amplitude
does not depend on the choice of cocycles (see also the discussion
in ref.\ \cite{PR4}). \par
Finally, notice that the ordering chosen for the fermions in
eq.\ (\ref{vertop}) is not accidental, it is a consequence of
the introduction of cocycles and of the fact that
in Minkowski space-time the number operator related
to fermion $\psi^{0}$ is anti-hermitian, rather than hermitian
as the other ones.
Therefore to retain the correct hermiticity properties
of the bosonized fermions, it is necessary to assign labels $33$
and $34$ to the fermion associated with the time direction in space-time
and to the superghost respectively \cite{PR3}.

\subsection{Gamma Matrices}
We define a set of four-dimensional gamma matrices by
means of the OPE between the real space-time fermions
$\psi^{\mu}$ and the spin
field $ S_{\bf A} \equiv  S_ A S^{(34)}_{a_{34}}$:
\be
 \psi^{\mu} (z) S_{\bf A} (w) = \frac{1} {\sqrt{2}}
(\Gamma^{\mu})_{\bf A}^{\ \bf B} \frac{S_{\bf B} (w)}{\sqrt{z-
w}}.\label{gammadef}
\ee
The explicit representation of
these matrices depends on the choice of the cocycle signs $Y$.
In terms of  the usual Pauli matrices  we have ($\sigma_0$ is the
$2 \times 2$ unit matrix)
\bea
\Gamma^0 &= & \left( -i \prod_{j=1}^{33} Y_{33,j}\right)
              \otimes _{\bar{l}=1}^{7}\sigma_3^{(\bar{l})} \otimes
               \sigma_3^{(\bar{17})}
              \otimes _{l=24,28,29}\sigma_3^{(l)}
              \otimes \sigma_3^{(32)} \otimes
              \sigma_2^{(33)} \otimes
               \sigma_0^{(34)}\nonumber\\
\Gamma^1 &= & \left( \prod_{j=1}^{33} Y_{33,j}\right)
              \otimes _{\bar{l}=1}^{7}\sigma_3^{(\bar{l})} \otimes
              \sigma_3^{(\bar{17})}
              \otimes _{l=24,28,29}\sigma_3^{(l)}
              \otimes \sigma_3^{(32)} \otimes
              \sigma_1^{(33)} \otimes
               \sigma_0^{(34)}\nonumber\\
\Gamma^2 &= & \left( - \prod_{j=1}^{33} Y_{32,j}\right)
              \otimes _{\bar{l}=1}^{7}\sigma_3^{(\bar{l})} \otimes
              \sigma_3^{(\bar{17})}
              \otimes _{l=24,28,29}\sigma_3^{(l)}
              \otimes \sigma_1^{(32)} \otimes
               \sigma_0^{(33)} \otimes
               \sigma_0^{(34)}\nonumber\\
\Gamma^3 &= & \left( - \prod_{j=1}^{33} Y_{32,j}\right)
              \otimes _{\bar{l}=1}^{7}\sigma_3^{(\bar{l})}
              \otimes \sigma_3^{(\bar{17})}
              \otimes _{l=24,28,29}\sigma_3^{(l)}
              \otimes \sigma_2^{(32)} \otimes
              \sigma_0^{(33)} \otimes
               \sigma_0^{(34)}.
\label{gammamatrix}
\eea
For a generic ground state eq.\ (\ref{vertop}) it is also convenient to
define a ``generalized charge conjugation matrix" ${\bf C}$ by
\be
S_{\bf A}(z,\bar{z}) S_{\bf B} (w,\bar{w}) \eqope
{\bf C}_{A B} \delta_{a_{34},b_{34}} \
\frac{1}{(z-w)^p}  \frac{1}{(\bar{z} - \bar{w})^{\bar{p}}},
 \ee
where $\bar{p}=\sum_{l=1}^{22} (\bar{a}_l)^2$ and
$p=\sum_{l=23}^{34} (a_l)^2$.
This matrix is related to the choice of cocycles by
\be
 {\bf C}_{A B} = e^{i \pi {\bf A} \cdot Y \cdot {\bf B} }
\left(\prod_{L=1}^{33}\delta_{A_L + B_L}\right)\delta_{a_{34},b_{34}} ,
\ee
where $a_{34} = b_{34}$ are given by eq.\ (\ref{supghcharge}).\par
In the computation of the amplitude it will be useful to
introduce also another set of gamma matrices, which are defined
by means of the OPE between the real space-time
fermions $\psi^{\mu}$ and the four-dimensional spin
field operator
$S_{\alpha} \equiv S^{(32)}_{a_{32}} S^{(33)}_{a_{33}}$:
\be
\psi^{\mu} (z) S_{\alpha} (w) \eqope {1 \over \sqrt{2}}
(\gamma^{\mu})_{\alpha}^{\ \beta} {S_{\beta} (w) \over
\sqrt{z-w} }.
\label{gammadef4}
\ee
As for the matrices in eq.\ (\ref{gammamatrix}), the explicit
representation of this new set of gamma matrices depends on the value
of $Y_{33,32}$
\bea
\gamma^0  &=& \left(iY_{33,32}\right)\sigma_3^{(32)} \otimes \sigma_1^{(33)}
\nonumber\\
\gamma^1 & = & \left(Y_{33,32}\right) \sigma_3^{(32)} \otimes\sigma_2^{(33)}
\nonumber\\
\gamma^2 & = & - \sigma_2^{(32)} \otimes \sigma_0^{(33)} \nonumber\\
\gamma^3 & = & \sigma_1^{(32)} \otimes \sigma_0^{(33)}.
\label{gammamatrix4}
\eea
\subsection{Vertex Operators in Our Model}
We are now ready to introduce the explicit vertex operators
necessary for the computation of the amplitude we are interested
in. The process we want to study is the Compton scattering of a
photon and a massless fermion.
The photons are the $U(1)$ gauge bosons belonging to the sector
${\bf W_0}$. The vertex operator describing a $U(1)$ gauge boson
with polarization $\epsilon$
and momentum $k$ in the superghost picture $q=-1$ is~\cite{FMS}
\be
{\cal V}^{(-1)}_{{\rm photon}} (z,\bar{z};k;\epsilon) = {\kappa
\over \pi} \bar\psi_{(\overline{17})}\bar\psi_{(\overline{17})}^*
(\bar{z}) \
\epsilon \cdot \psi (z) \, e^{-\phi(z)} \, (c_{(34)})^{-1}
e^{ik\cdot
X(z,\bar{z})}, \label{photon0}
\ee
where the gravitational coupling $\kappa$ is related to Newton's
constant by
$\kappa^2 = 8\pi G_N$, $\epsilon \cdot \epsilon = 1$,
$k^2 = \epsilon \cdot k = 0$.
The picture changed version of the same vertex is
\be
{\cal V}^{(0)}_{{\rm photon}} (z,\bar{z};k;\epsilon) =
-i {\kappa \over \pi} \bar\psi_{(
\overline{17})}\bar\psi_{(\overline{17})}^* (\bar{z}) \
\left[
\epsilon \cdot \partial_z X(z) - i k\cdot \psi (z) \epsilon \cdot
\psi (z)
\right] \,  e^{ik\cdot X(z,\bar{z})} \ . \label{photon}
\ee
As space-time fermions, we choose the massless fermions which
form the ground state of the
sector ${\bf W}_{134}$. They have $U(1)$ charge $\pm 1/2$ and they are
described by the vertex operator
\be
{\cal V}^{(-1/2)} (z,\bar{z};k;{\bf V})=
 \frac{\kappa}{\pi} \ {\bf V}^A
S_A (z,\bar{z}) \
e^{-\frac12\phi(z)} (c_{(34)})^{-1/2}\ e^{ik\cdot
X(z,\bar{z})},\label{fermion}
\ee
where $k$ is the momentum and  the label $-1/2$ indicates the
superghost vacuum charge.
$S_A$ is the spin field which creates the Ramond ground state from
the conformal vacuum
\be
S_A \ = \
\left(\prod_{l=1}^7 \bar{S}^{(\bar{l})}_{\bar{a}_l}
(\bar{z})\right) \
\bar{S}^{(\bar{17})}_{\bar{a}_{17}} (\bar{z})
\left(\prod_{l=24,28,29} S^{(l)}_{a_l}(z)\right) \
S^{(l)}_{\alpha}(z).
\ee
The ``spinor'' ${\bf V}$ decomposes accordingly
\be
{\bf V}^A = \bar{V}_{SO(14)}^{\bar{a}_1,\dots,\bar{a}_7} \
\bar{V}^{\bar{a}_{17}}_{U(1)} \ \left(\prod_{l=24,28,29}
v_{(l)}^{a_l}\right)
\ V^{\alpha}.
\ee
The left-moving spinor indices $\bar{a} =
(\bar{a}_1,\ldots,\bar{a}_7)$
all take values $\pm 1/2$ and indicate that the fermion transforms
in the spinor representation of the first $SO(14)$.
$\bar{a}_{17} = \pm \frac12$ is the $U(1)$ charge.
The right-moving spinor indices also take values $\pm 1/2$:
$\alpha= (a_{32}, a_{33})$
is the four-dimensional space-time spinor index, while the
others are just enumerative family indices.\par
In order to describe physical external states, the vertex operators
$\bar{c} c {\cal V}$, with ${\cal V}$ as in eqs.\ (\ref{photon}) and
(\ref{fermion}), must be BRST invariant \cite{FMS,PR1}.
The BRST currents are given by
\bea
j_{\rm BRST} & = & c T_B^{[X,\psi,\beta,\gamma]} - c b
\partial c - T_F^{[X,\psi]} e^{\phi} c_{(34)} \, \eta - {1 \over 4}
e^{2\phi} (c_{(34)})^2 \, \eta (\partial \eta) b \nonumber\\
\bar{j}_{\rm BRST} & = & \bar{c} \bar{T}_B^{[\bar{X},\bar{\psi}]} - \bar{c}
\bar{b} \bar{\partial} \bar{c},  \label{BRST}
\eea
where $T_B$ and $\bar{T}_B$ are the energy-momentum tensors.
The first-order pole in the OPE of $\bar{j}_{\rm BRST}$ with $\bar{c} c
{\cal V}$, as well as the first order pole in the OPE of the first two
terms of $j_{\rm BRST}$ with $\bar{c} c {\cal V}$, vanish merely by
imposing that the vertex operator ${\cal V}$ is a primary
conformal field of dimension one.
The last term in $j_{\rm BRST}$ has a non-singular OPE with
$\bar{c} c {\cal V}$ for any  operator ${\cal V}$ whose
superghost part is given by $e^{-\phi}$ or $e^{-\phi/2}$. Therefore the
BRST-invariance is reduced to the requirement that the first-order pole
in the OPE
$e^{\phi(w)} c_{(34)} T_F^{[X,\psi]}(w) \ c\bar{c} {\cal V}(z,\bar{z})$
should vanish.
For a gauge boson, this equation becomes the transversality condition
\be
\epsilon\cdot k \ =\ 0\ ,
\ee
whereas for the space-time fermion it becomes the
``Dirac equation'' \cite{PR1}
\be
{\bf V}^T (k) \Gamma^{\mu}k_{\mu}=0 \label{Diraceq},
\ee
with $\Gamma^{\mu}$ given by eqs.\ (\ref{gammamatrix}).

\subsection{Incoming and Outgoing Vertex Operators}
In the formula for the 1-loop $T$-matrix element eq.\  (\ref{Tmatrix})
we have quoted an explicit value for the normalization coefficient
$C_{g=1}$. Obviously this value is meaningful only when one specifies
at the same time also the normalization of all other ingredients of
the formula. In particular we need to normalize consistently all the
vertex operators.
The general issue can be briefly summarized as follows.
In string theory we compute the connected part of a transition amplitude
$<\lambda_1, \ldots;in\vert S\vert \ldots,\lambda_N;in>$ by the ``master''
formula eq.\ (\ref{Smatrix}) (and eq.\  (\ref{Tmatrix})), where each
single-string state ---whether appearing as a bra or a ket---
is represented by a vertex operator ${\cal W}$.
Now suppose we are given a vertex operator
${\cal W}_{\vert\lambda;in>}$ representing a single-string ket-state, what
is the vertex operator ${\cal W}_{<\lambda;in\vert}$ representing the same
single-string state but now as a bra? The problem of course has to do
with the hermiticity and unitarity properties of the scattering amplitude,
and must be solved at the same time as solving the problem of the
normalization of the vertex operators. The general discussion can be found
in ref.\ \cite{PR2}, here we just report the final result for the
vertex operators of interest to us. First of all, for the photons the
bra and ket vertex operators are identical and normalized
as in eqs.\ (\ref{photon0}) and (\ref{photon}).

The situation is different for the fermion vertex operator in eq.\
(\ref{fermion}). Let us assume that this vertex operator describes
a ket-state. Then the analysis of ref.\ \cite{PR2} implies that
the ``spinor'' ${\bf V}^A$ satisfies the following normalization condition
\be
{\bf V}^\dagger (k) {\bf V}(k)\ =\ \sqrt{2} \vert k^0\vert\ .
\ee
Moreover if we denote the vertex operator describing the same state
but outgoing by
\be
{\cal V}^{(-1/2)} (z,\bar{z};k;{\bf W})=
 \frac{\kappa}{\pi} \ {\bf W}^A
S_A (z,\bar{z}) \
e^{-\frac12\phi(z)} (c_{(34)})^{-1/2}\ e^{-ik\cdot
X(z,\bar{z})},\label{fermionout}
\ee
then the vector ${\bf W}^A$ is related to the ``spinor'' ${\bf V}^A$
now describing the outgoing fermion by
\be
{\bf W}^T\ =\ -i Y_{33,34} {\bf V}^\dagger\sigma_1^{(33)} {\bf C}\ .
\ee
An explicit example of how these vertex operators are used in the
computation of a scattering amplitude can be found in section
8 of ref.\ \cite{PR2} (see also section 2 of ref.\ \cite{PPlett}).\par
Finally, the gauge coupling constant $e$ is expressed
in terms of the constant $\kappa$ by the relation
$$
e^2=\kappa^2/(2 \alpha')
$$
as it has been described in ref.\ \cite{PPlett} by the comparison of the
tree level (genus zero) Compton scattering amplitude and the field
theory result.

\section{The Explicit Computation of the Amplitude}
\setcounter{equation}{0}
Having introduced all the ingredients, we can now describe the
explicit computation of the 1-loop Compton scattering amplitude
in our chosen four-dimensional heterotic string model.
Being a ``Compton'' scattering amplitude, the incoming/outgoing states are
a photon and an ``electron'' (or ``positron''), which in a string model
are represented by a massless space-time fermionic state charged
under the $U(1)$
component of the gauge group. Notice that in string theory we usually
cannot find a state which is {\it only\/} charged under the $U(1)$,
and indeed our fermion is also charged under the first $SO(14)$ and
carries some other enumerative indices, as discussed in the
previous section.

We will label the incoming particles by the indices 2 and 4, and the
outgoing with 1 and 3. Moreover we choose all momenta as incoming
(not as in eq.\ (\ref{fermionout})). For what concerns the $N_{B}+N_{FP}=3$
PCO's, we insert one at each of the photon vertex operators, changing
them from the $q=-1$ to the $q=0$ picture. Thus we remain with one
PCO inserted at an arbitrary point $w$ and the fermionic vertex operators
are in the $q=-1/2$ picture. This choice is the most convenient from the
following point of view. First of all, it is more convenient to have
the photon vertex operators in the zero picture than to have to deal with
two other PCO's inserted at arbitrary points on the world-sheet;
in this way the number of terms to be computed decreases and no
complication is added. On the
other hand, it is technically not convenient to absorb the last
PCO in the vertex operator of a fermion since this will then be
in the $+1/2$ picture, and the expressions to be computed will
become much more cumbersome. Moreover, having one free PCO on the
world-sheet will give us a powerful tool for checking the correctness
of the computation, since the final scattering amplitude must be
independent on the position of the insertion of any PCO.

The form of the amplitude which we start from is then
\bea
T_{g=1}&(& e^{\pm} \ + \ \gamma  \ \rightarrow \
e^{\pm} \
+ \ \gamma )= \nonumber\\
   &C&_{g=1} \int {\rm d}^2\tau {\rm d}^2z_2 {\rm d}^2z_3
      {\rm d}^2z_4 \
      \sum_{m_i,n_j} C^{\bbf {\alpha}}_{\bbf {\beta}} \
      \wew{\vert (\eta_{\tau}\vert b)
(\eta_{z_2}\vert b)
      (\eta_{z_3}\vert b)(\eta_{z_4}\vert b)
\nonumber\\
  &\times&  c(z_1)c(z_2)c(z_3)c(z_4)\vert^2 \
      \Pi(w) \ {\cal V}^{(0)}_{\rm photon}
      (z_1,\bar{z}_1;k_1;\epsilon_1)
      {\cal V}^{(0)}_{\rm photon}
(z_2,\bar{z}_2;k_2;\epsilon_2)\nonumber\\
  &\times & {\cal V}^{(-1/2)} (z_3,\bar{z}_3;k_3;{\bf
W}_3) \
      {\cal V}^{(-1/2)} (z_4,\bar{z}_4;k_4;{\bf V}_4)},
\label{oneloop1}
\eea
where we used the translation invariance of the torus to fix the
position $z_1$ of the outgoing photon vertex operator to an
arbitrary value.

If we substitute in eq.(\ref{oneloop1}) the explicit expression for the PCO
(eq.(\ref{pco})) and the vertex operators (eqs.\
(\ref{photon}) and  (\ref{fermion})), we obtain
\bea
T^{1-{\rm loop}}&=&
  -i \left( \frac{\kappa^2}{4\pi^3\alpha'} \right)^2  {\bf
W}^{\bf A}_3{\bf V}^{\bf B}_4
\int {\rm d}^2\tau  {\rm d}^2z_2 {\rm d}^2z_3 {\rm d}^2z_4
\sum_{m_i,n_j} C^{\bbf{\alpha}}_{\bbf{\beta}}
e^{i\pi {\bf A} \cdot Y\cdot {\bf B}} \nonumber\\
& &
\wew{\vert (\eta_{\tau}\vert b)
(\eta_{z_2}\vert b)(\eta_{z_3}\vert b)(\eta_{z_4}\vert b)
c(z_1)c(z_2)c(z_3)c(z_4)\vert^2 }
\times T_{L}
\times T_R,
\label{fundforc}
\eea
where
\be
T_{L} \ =\  \wew{
\prod_{l=1}^{7}
\left(\bar{S}_{\bar{a}_l}^{(\bar{3})} (\bar{z}_3)
\bar{S}_{\bar{b}_l}^{(\bar{l})} (\bar{z}_4)\right) \
\bar{\partial}\bar{\phi}_{(\overline{17})}(\bar{z_1})
\bar{\partial}\bar{\phi}_{(\overline{17})}(\bar{z_2})
\bar{S}_{\bar{a}_{17}}^{(\overline{17})} (\bar{z}_3)
\bar{S}_{\bar{b}_{17}}^{(\overline{17})}
(\bar{z}_4)} \label{LmP}
\ee
and
\bea
T_R\ = & & \wew{
\prod_{l=24,28,29,32,33} \left(S_{a_l}^{(l)}
(z_3) S_{b_l}^{(l)}(z_4) \right)
\left[\epsilon_1\cdot \partial X(z_1) -i k_1 \cdot
\psi(z_1) \epsilon_1 \cdot \psi(z_1)\right] \times \nonumber\\
 & &\left[\epsilon_2 \cdot \partial X(z_2) -i k_2 \cdot
\psi(z_2) \epsilon_2 \cdot \psi(z_2) \right] \
\left(\partial X (w) \cdot \psi(w) \right) \times
\nonumber\\
& &  e^{ik_3 \cdot X(z_3,\bar{z}_3)} e^{ik_4 \cdot
X(z_4,\bar{z}_4)}
e^{ik_1\cdot X(z_1,\bar{z}_1)} e^{ik_2\cdot
X(z_2,\bar{z}_2)}
e^{-\frac12\phi(z_3)} e^{-\frac12\phi(z_4)} e^{\phi(w)}}.
\label{RmP}
\eea
In $T_L$ and $T_R$, we rearranged the operators in such a way to group all
fermionic right-movers and all fermionic left-movers together
(but notice that $T_R$ depends on $\bar{z}$ due to the presence
of the $X_\mu$).
To do this we moved the fermions $\psi^{\mu}$ across the
superghosts, and the spin fields across each other. In doing so,
the cocycles give rise to phases which combine in the
overall factor $exp \{i\pi {\bf A} \cdot Y\cdot {\bf
B}\}$ in eq.(\ref{fundforc})~\cite{PR1}.
Notice also that,  because of  fermion number
conservation, only the first term of the PCO operator
gives a non-zero contribution to the correlation function, so
that effectively
$\Pi \simeq  -i e^\phi c_{(34)}\partial X \cdot \psi$.
\subsection{Computation of Correlators}
We now turn our attention to the computation of the correlators appearing
in eqs.\ (\ref{LmP}) and  (\ref{RmP}). The correlators involving the
bosonic coordinate fields $X^\mu$ can be computed using the Wick theorem
and  the contraction given by the bosonic Green function (see Appendix A)
and will not be reported explicitly here.
We instead give the correlators involving the world-sheet
fermions, the spin fields, the ghosts and superghosts.
As already mentioned, the notation $\langle \langle
\dots  \rangle \rangle$ indicates correlators including their
complete contribution to the partition function. For each complex
fermion, it is convenient to introduce a
correlation function $\langle \dots \rangle$ where the
non-zero mode part of the
partition function has been removed:
\be
\wew{{\cal O}_1(z_1) \dots {\cal O}_N(z_N)}_{(l)}\ =\
\prod_{n=1}^\infty (1-k^n)^{-1} \langle {\cal O}_1(z_1)
\dots {\cal O}_N
(z_N) \rangle _{(l)}.\label{gencorr}
\ee
The subscript $(l)$ indicates that the correlator
depends on the spin structure. Moreover, notice that
\be
\langle {\bf 1} \rangle_{(l)} = \Theta \left[
{}^{\alpha_l}_{\beta_l}
\right] (0 \vert \tau) ,
\ee
which vanishes when the spin structure is odd.

All the correlators involving spin fields are obtained from
the fundamental one $\langle \prod_{i=1}^N
e^{q_i\phi(z_i)}\rangle$ (given in Appendix A), and
correlators involving $\partial\phi$
can be obtained from this by differentiation.
The basic spin field correlator is
\be
\vev{S_{a_l}^{(l)} (z_3) S_{b_l}^{(l)} (z_4) }_{(l)} \ =\
\left( (\sigma_3^{(l)})^{S_l} \sigma_1^{(l)}\right)_{a_l,b_l}
\vev{S_+(z_3) S_-(z_4)}_{(l)},
\label{spincorr}
\ee
where we introduced
\be
S_l\equiv (1-2\alpha_l)(1+2\beta_l),
\ee
which is $0 (1) {\rm MOD2} $ depending on whether the spin structure
$\left[{}^{\alpha_l}_{\beta_l}\right]$ is even (odd). Notice that the
correlator (\ref{spincorr}) develops a dependence on the sign of the charge
$a_l$ whenever the spin structure is odd. The correlator $\vev{ S_+
(z_3) S_- (z_4)}$ is given by
\bea
\vev{S_+(z_3) S_-(z_4)}& = &
\vev{e^{\frac12\phi(z_3)}e^{-\frac12\phi(z_4)}} \nonumber\\
& = & \left( E(z_3,z_4)\right)^{-\frac14}
\Teta\alpha\beta(\frac12\nu_{34} \vert\tau).
\eea

The other spin field correlators in equations (\ref{LmP}) and
(\ref{RmP}) are
\bea
&\langle\bar\partial\bar\phi_{(\overline{17})}&(\bar{z}_1)
\bar\partial\bar\phi_{(\overline{17})}(\bar{z}_2) \,
\bar{S}_{\bar{a}_{17}}^{(\overline{17})}(\bar{z}_1) \,
\bar{S}_{\bar{b}_{17}}^{(\overline{17})}(\bar{z}_2)\rangle  =
\label{correl1}\\
& & \left( (\sigma_3^{(\overline{17})})^{\overline{S}_{17}}
\sigma_1^{(\overline{17})}\right)_{\bar{a}_{17}\bar{b}_{17}} \,
\vev{\bar{S}_+(\bar{z}_1)\bar{S}_-(\bar{z}_2)}_{(\overline{17})}\nonumber\\
& & \times \left\{ \partial_{\bar{z}_1}
\partial_{\bar{z}_2} \log \bar{E}(\bar{z}_1,\bar{z}_2)
+\frac14 \ \partial_{\bar{z}_1}\log \frac{\bar{E}(\bar{z}_1,\bar{z}_3)}
{\bar{E}(\bar{z}_1,\bar{z}_4)} \
 \partial_{\bar{z}_2}\log \frac{\bar{E}(\bar{z}_2,\bar{z}_3)}
{\bar{E}(\bar{z}_2,\bar{z}_4)} \right. \nonumber\\
                &  & \quad + \frac12 \frac{\bar{\omega}(\bar{z}_1)}{2\pi i} \
\partial_{\bar{z}_2}\log \frac{\bar{E}(\bar{z}_2,\bar{z}_3)}
{\bar{E}(\bar{z}_2,\bar{z}_4)}\partial_{\nu} \log \bar{\Theta}
\left[{}^{\bar{\alpha}_{17}}_{\bar{\beta}_{17}} \right]
(\nu\vert\bar{\tau})\vert_{\nu=\frac12\bar{\nu}_{34}} \nonumber\\
                &  & \quad + \frac12
\frac{\bar{\omega}(\bar{z}_2)}{2\pi i} \
\partial_{\bar{z}_1}\log \frac{\bar{E}(\bar{z}_1,\bar{z}_3)}
{\bar{E}(\bar{z}_1,\bar{z}_4)} \partial_{\nu} \log \bar{\Theta}
\left[{}^{\bar{\alpha}_{17}}_{\bar{\beta}_{17}} \right]
(\nu\vert\bar{\tau})\vert_{\nu=\frac12\bar{\nu}_{34}} \nonumber\\
                &  & \quad + \left. \frac{\bar{\omega}(\bar{z}_1)}{2\pi i}
\frac{\bar{\omega}(\bar{z}_2)}{2\pi i} \
\left(\bar{\Theta} \left[{}^{\bar{\alpha}_{17}}_{\bar{\beta}_{17}} \right]
(\frac12\bar{\nu}_{34}\vert\bar{\tau}) \right)^{-1} \ \partial^2_{\nu}
\bar{\Theta}
(\nu\vert\bar{\tau})\vert_{\nu=\frac12\bar{\nu}_{34}} \right\},\nonumber\\
& \langle  S_{a_{32}}^{(32)} &(z_3)  S_{b_{32}}^{(32)}(z_4)\,
S_{a_{33}}^{(33)}(z_3)\, S_{b_{33}}^{(33)} (z_4)\psi^\tau (w) \rangle
= \label{correl2}\\
& &  {1\over \sqrt{2} }
\left(\vev{S_+(z_3)S_-(z_4)}_{(32)} \right)^2\,
\Ical{\alpha_{32}}{\beta_{32}}{(z_3,z_4,w)}\
(\gamma^\tau (\gamma^5)^{S} {\rm C})_{\alpha\beta} \,
e^{-i\pi a_{33}Y_{33,32}b_{32}}, \nonumber\\
&\langle S_{a_{32}}^{(32)}& (z_3)  S_{b_{32}}^{(32)}(z_4)\,
S_{a_{33}}^{(33)}(z_3)\, S_{b_{33}}^{(33)} (z_4)
\psi^\mu\psi^\nu(z_1) \, \psi^\tau (w)\rangle= \label{correl3}\\
           & & -{1\over \sqrt{2}}
\left(\vev{S_+(z_3)S_-(z_4)}_{(32)} \right)^2\,
\Ical{\alpha_{32}}{\beta_{32}}{(z_1,z_3,z_4)}\nonumber\\
 & & \times
\left\{\GGm{\alpha_{32}}{\beta_{32}}{(z_1,z_3,z_4,w)}\,
(\gamma^{\mu\nu\tau}(\gamma^{5})^{S}{\rm C})_{\alpha\beta} \right.
\nonumber\\
          & & \quad \left.+ \GGp{\alpha_{32}}{\beta_{32}}{(z_1,z_3,z_4,w)}\,
\left((g^{\mu\tau}\gamma^\nu -g^{\nu\tau}\gamma^\mu)
(\gamma^{5})^{S}{\rm C}\right)_{\alpha\beta} \right\}
e^{-i\pi a_{33}Y_{33,32}b_{32}}, \nonumber\\
&\langle  S_{a_{32}}^{(32)} & (z_3)  S_{b_{32}}^{(32)}(z_4)\,
S_{a_{33}}^{(33)}(z_3)\, S_{b_{33}}^{(33)} (z_4)
\psi^\mu\psi^\nu(z_1) \,\psi^\rho\psi^\sigma(z_2) \,
\psi^\tau (w) \rangle= \label{correl4}\\
          & & {1\over \sqrt{2}}
\left(\vev{S_+(z_3)S_-(z_4)}_{(32)} \right)^2
          \left\{-\Ical{\alpha_{32}}{\beta_{32}}
          {(z_3,z_4,w)}\right.\nonumber\\
& &\quad \times \left(\GGp{\alpha_{32}}{\beta_{32}}{(z_1,z_2,z_3,z_4)}\right)^2
\left((g^{\mu\rho}g^{\nu\sigma}-g^{\mu\sigma}g^{\nu\rho}) \gamma^\tau
(\gamma^{5})^{S}{\rm C}\right)_{\alpha\beta} \nonumber\\
          & &\quad + \Ical{\alpha_{32}}{\beta_{32}}{(z_1,z_3,z_4)}\,
\GGp{\alpha_{32}}{\beta_{32}}{(z_2,z_3,z_4,w)}\,
\GGp{\alpha_{32}}{\beta_{32}}{(z_2,z_1,z_3,z_4)} \nonumber\\
          & & \qquad\times
\left[\left((g^{\nu\sigma}g^{\rho\tau}-g^{\nu\rho}g^{\sigma\tau})
\gamma^\mu -
(g^{\mu\sigma}g^{\rho\tau}-g^{\mu\rho}g^{\sigma\tau})
\gamma^\nu \right)
(\gamma^{5})^{S}{\rm C}\right]_{\alpha\beta} \nonumber\\
          & & \quad +
\Ical{\alpha_{32}}{\beta_{32}}{(z_2,z_3,z_4)}\,
\GGp{\alpha_{32}}{\beta_{32}}{(z_1,z_3,z_4,w)}\,
\GGp{\alpha_{32}}{\beta_{32}}{(z_1,z_2,z_3,z_4)} \nonumber\\
          & & \qquad \times
\left[\left((g^{\nu\sigma}g^{\mu\tau}-g^{\mu\sigma}g^{\nu\tau})
\gamma^\rho -
(g^{\mu\tau}g^{\nu\rho}-g^{\mu\rho}g^{\nu\tau})
\gamma^\sigma \right)
(\gamma^{5})^{S}{\rm C}\right]_{\alpha\beta} \nonumber\\
          & & \quad+
\Ical{\alpha_{32}}{\beta_{32}}{(z_3,z_4,w)}\,
\GGp{\alpha_{32}}{\beta_{32}}{(z_1,z_2,z_3,z_4)}\,
\GGm{\alpha_{32}}{\beta_{32}}{(z_1,z_2,z_3,z_4)} \nonumber\\
          & & \qquad\times
\left((g^{\nu\sigma}\gamma^{\mu\rho\tau}
-g^{\nu\rho}\gamma^{\mu\sigma\tau}
+g^{\mu\rho}\gamma^{\nu\sigma\tau}
-g^{\mu\sigma}\gamma^{\nu\rho\tau})
(\gamma^{5})^{S}{\rm C}\right)_{\alpha\beta}  \nonumber\\
          & & \quad +
\Ical{\alpha_{32}}{\beta_{32}}{(z_1,z_3,z_4)}\,
\GGp{\alpha_{32}}{\beta_{32}}{(z_2,z_3,z_4,w)}\,
\GGm{\alpha_{32}}{\beta_{32}}{(z_1,z_2,z_3,z_4)}\nonumber\\
          & & \qquad\times \left((g^{\rho\tau}\gamma^{\mu\nu\sigma}
-g^{\sigma\tau}\gamma^{\mu\nu\rho})
(\gamma^{5})^{S}{\rm C}\right)_{\alpha\beta} \nonumber\\
         & & \quad+
\Ical{\alpha_{32}}{\beta_{32}}{(z_2,z_3,z_4)}\,
\GGp{\alpha_{32}}{\beta_{32}}{(z_1,z_3,z_4,w)}\,
\GGm{\alpha_{32}}{\beta_{32}}{(z_1,z_2,z_3,z_4)}\nonumber\\
         & & \qquad\left. \times \left((g^{\mu\tau}\gamma^{\nu\rho\sigma}
-g^{\nu\tau}\gamma^{\mu\rho\sigma})
(\gamma^{5})^{S}{\rm C}\right)_{\alpha\beta} \right\}
e^{-i\pi a_{33}Y_{33,32}b_{32}},\nonumber
\eea
In these equations $\gamma^\mu$ and $\gamma^{\mu\nu\rho}$ are the
four-dimensional gamma matrices defined in eq.\ (\ref{gammadef4}) and their
antisymmetrized products.
We also introduced the shorthand notations $S=S_{32}$ and
\be
{\rm C}_{\alpha \beta} \equiv \delta_{a_{32}+b_{32}}
\delta_{a_{33}+b_{33}} e^{i \pi a_{33} Y_{33,32}
b_{32}} \label{abbr},
\ee
and defined the function of the world-sheet coordinates
\bea
&{\cal I}&\left[{}^{\alpha}_{\beta} \right] (z;z_3,z_4)  =
\sqrt{\frac{E(z_3,z_4)} {E(z,z_3)E(z,z_4)}}
\frac{\Theta \left[{}^{\alpha}_{\beta} \right] (\mu_z \vert \tau)}
{\Theta \left[ {}^{\alpha}_{\beta} \right]
(\frac12 \nu_{34} \vert\tau)} \nonumber\\
&I& \left[{}^{\alpha}_{\beta} \right] (z;z_3,z_4) =
\partial_z \log \frac{E(z,z_3)}{E(z,z_4)}
+ 2 \frac{\omega(z)}{2\pi i}
\partial_{\nu} \log {\Theta \left[{}^{\alpha}_{\beta} \right] (\nu
\vert \tau)}
\vert_{\nu=\frac12\nu_{34}}, \nonumber\\
& G^{+}&\left[{}^{\alpha}_{\beta} \right] (z,w;z_3,z_4)  =
\frac{1}{2 E(z,w)}
\left\{\frac{\Theta \left[{}^{\alpha}_{\beta}
\right](\rho_{z,w}\vert\tau)}
{\Theta \left[{}^{\alpha}_{\beta}
\right](\frac12\nu_{34}\vert\tau)}
\sqrt{\frac{E(z,z_3) E(w,z_4)}{E(w,z_3) E(z,z_4)}} \right. \nonumber\\
 & &   \qquad\quad \left. + \ \frac{\Theta
\left[{}^{\alpha}_{\beta} \right]
(\rho_{w,z}\vert\tau)} {\Theta \left[{}^{\alpha}_{\beta} \right]
(\frac12\nu_{34}\vert\tau)}
\sqrt{\frac{E(w,z_3) E(z,z_4)}{E(z,z_3) E(w,z_4)}} \right\} \nonumber\\
 & & \quad = \frac{\Ical\alpha\beta(w,z_3,z_4)}{\Ical\alpha\beta
(z,z_3,z_4)} \left\{\partial_z\log \frac{E(z,w)}{\sqrt{E(z,z_3)E(z,z_4)}}
\right. \nonumber\\
 & & \qquad \quad \left.+
{\omega(z)\over 2\pi i} \partial_{\nu}\log\Teta\alpha\beta(\nu\vert\tau)
\vert_{\nu=\mu_w} \right\}, \nonumber\\
& G^{-}&\left[{}^{\alpha}_{\beta} \right] (z,w;z_3,z_4)  =
\frac{1}{2 E(z,w)}
\left\{\frac{\Theta \left[{}^{\alpha}_{\beta}
\right](\rho_{z,w}\vert\tau)}
{\Theta \left[{}^{\alpha}_{\beta}
\right](\frac12\nu_{34}\vert\tau)}
\sqrt{\frac{E(z,z_3) E(w,z_4)}{E(w,z_3) E(z,z_4)}} \right. \nonumber\\
 & &   \qquad \quad \left. - \ \frac{\Theta
\left[{}^{\alpha}_{\beta} \right]
(\rho_{w,z}\vert\tau)} {\Theta \left[{}^{\alpha}_{\beta} \right]
(\frac12\nu_{34}\vert\tau)}
\sqrt{\frac{E(w,z_3) E(z,z_4)}{E(z,z_3) E(w,z_4)}} \right\} \nonumber\\
& &\quad = {1\over 2}\Ical\alpha\beta(z,z_3,z_4)\Ical\alpha\beta
(w,z_3,z_4),
\label{eIIGG}
\eea
where
\bea
& \nu &_{12} \equiv \int^{z_3}_{z_4} \frac\omega{2\pi
i} \nonumber\\
& \mu &_z \equiv \int^z \frac\omega{2\pi i} -\frac12
\int^{z_3}
\frac\omega{2\pi i} -\frac12 \int^{z_4}\frac\omega{2\pi
i}  \nonumber\\
& \rho &_{z,w} \equiv \int^z_w \frac\omega{2\pi i} +
\frac12 \int^{z_3}_{z_4}\frac\omega{2\pi i}.
\eea
We do not show here the steps leading to expressions
(\ref{correl1}-\ref{correl4}), but we refer to~\cite{PR1}, where
an explicit example of this kind of computation has been given.
The crucial point is the recovering of the explicit Lorentz covariance
of the correlators, which is lost once the fermions have been
bosonized.
First of all it is necessary to reconstruct the four-dimensional gamma
matrix algebra out of the phases  coming from the fundamental
correlator defined in Appendix A (eq.(\ref{aboscorr})). Moreover
it turns out that
correlators (\ref{correl3}) and (\ref{correl4}) involving several
space-time fermions $\psi^{\mu}$ have
different expressions in terms of theta-functions for different values
of the Lorentz indices.
For instance, the two different expressions we give for the functions
$G^{\pm}\left[ {}^{\alpha}_{\beta} \right]$ appear
when we compute the correlator (\ref{correl3}) for different values of
the indices $\mu,\nu,\tau$.
These two expressions, as well as similar ones coming from the other
correlators, must be
identical (up to different phases due to the gamma matrices),
otherwise the amplitude would not be Lorentz
covariant. It is therefore necessary to prove some
identities involving theta-functions, mostly of the form of the
trisecantic identity ~\cite{Fay}.
In Appendix B we give the list of the identities we needed to prove in
the case of our four particle amplitude, while in Appendix E of
ref.~\cite{PR1} it is sketched the proof of one of these identities.\par
Finally there are the correlators for the reparametrization ghosts and
the superghosts. They can be computed using the general expressions eqs.\
(\ref{bone}), (\ref{bseven}) of Appendix A  and are given by
\bea
& \langle\langle &e^{-\frac12\phi(z_3)}  e^{-\frac12\phi(z_4)}
e^{\phi(w)} \rangle\rangle =
 (-1)^{S_{32}} k^{1/2} \prod_{n=1}^{\infty} (1-k^n) \label{correl5}\\
 & & \quad \times{(\omega(z_3) \omega(z_4))^{1/2} \over \omega(w)}
\left(\vev{S_+(z_3)
S_-(z_4)}_{(32)}
\Ical{\alpha_{32}}{\beta_{32}}{(z_3,z_4,w)}\right)^{-1},\nonumber\\
& \langle\langle & \vert (\eta_{\tau}\vert b)(\eta_{z_2}\vert b)
(\eta_{z_3}\vert b)(\eta_{z_4}\vert b)
c(z_1)c(z_2)c(z_3)c(z_4)\vert^2 \rangle\rangle =
\left| {1 \over \omega(z_1)} \right|^2 \prod_{n=1}^{\infty}
\left| 1 - k^n \right|^4 \label{correl6}
\eea
\subsection{Using the GSO Projections}
The next step in the computation of the amplitude, after
substituting all the correlators into eq.\ (\ref{fundforc}),
is to simplify the factors of gamma and sigma matrices
appearing in the equation.
Indeed in each term of the amplitude there appears
a factor of the form
\bea
e^{i \pi {\bf A} \cdot Y \cdot {\bf B}}& &
\prod_{l=1}^7
 \left( (\sigma_3^{(\bar{l})})^{\bar{S}_l}
\sigma_1^{(\bar{l})} \right)_{\bar{a}_l \bar{b}_l}
\left( (\sigma_3^{(\overline{17})})^{\bar{S}_{17}}
\sigma_1^{(\overline{17})} \right)_{\bar{a}_{17}\bar{b}_{17}}\times\nonumber\\
& & \prod_{l=24,28,29}
\left( (\sigma_3^{(l)})^{S_l}
\sigma_1^{(l)} \right)_{a_l b_l}
 \left(\gamma_{*} \left( \gamma^5 \right)^{S_{32}}
\tilde{\rm C} \right)_{\alpha \beta}
e^{-i \pi a_{33} Y_{33,32}b_{32}}, \label{gammafactor}
\eea
where $\gamma_*$ denotes either $\gamma^{\mu}$ or
$\gamma^{\mu \nu \rho}$.
This product can be rearranged in such a way to reconstruct the
charge conjugation matrix ${\bf C}_{A B}$ and
the complete gamma matrices  (eq.\ (\ref {gammamatrix})).
This can be accomplished by moving all the
$\sigma_1^{(l)}$ to the left and by rewriting  the
phase factor in (\ref{gammafactor}) as
$e^{i\pi {\bf A} \cdot Y \cdot {\bf B}} = e^{i \pi
{\bf B}' \cdot Y
\cdot {\bf B}} e^{i \pi ({\bf A} - {\bf B}') \cdot Y
\cdot {\bf B}}$.
The first factor in this expression, together with $\sigma_1^{(l)}$,  are
what we need to reconstruct the
matrix ${\bf C}_{A B}$, while the second can be rewritten as a product
of $\sigma_3$
matrices acting directly on the spinor ${\bf W}_3$ and
contributes to give the complete gamma matrices.
In this way we obtain the following relation:
\bea
e^{i \pi {\bf A} \cdot Y \cdot {\bf B}}& &
\prod_{l=1}^7
 \left( (\sigma_3^{(\bar{l})})^{\bar{S}_l}
\sigma_1^{(\bar{l})} \right)_{\bar{a}_l \bar{b}_l}
\left( (\sigma_3^{(\overline{17})})^{\bar{S}_{17}}
\sigma_1^{(\overline{17})} \right)_{\bar{a}_{17}\bar{b}_{17}}\times\nonumber\\
& & \prod_{l=24,28,29}
\left( (\sigma_3^{(l)})^{S_l}
\sigma_1^{(l)} \right)_{a_l b_l}
 \left(\gamma_{*} \left( \gamma^5 \right)^{S_{32}}
\tilde{\rm C} \right)_{\alpha \beta}
e^{-i \pi a_{33} Y_{33,32}b_{32}} \ =  \nonumber\\
 & & \left( \prod_{l=1}^7
 (\sigma_3^{(\bar{l})})^{\bar{S}_l}
 (\sigma_3^{(\overline{17})})^{\bar{S}_{17}}
\prod_{l=24,28,29} (\sigma_3^{(l)})^{S_l}
\Gamma_* \left({\bf \Gamma}^5\right)^{S_{32}} {\bf C} \right)_{AB},
\label{cocgyma}
\eea
with $\Gamma_*$ now denoting either $\Gamma^{\mu}$ or
$\Gamma^{\mu \nu \rho}$.
Since these structures are sandwiched between the spinors ${\bf W}_3$
and ${\bf V}_4$, we can use the GSO projection conditions
for the sector ${\bf W}_{134}$, which these spinors belong to,
to further simplify expression (\ref{cocgyma}):
\be
\left( \prod_{l=1}^7
 (\sigma_3^{(\bar{l})})^{\bar{S}_l}
 (\sigma_3^{(\overline{17})})^{\bar{S}_{17}}
\prod_{l=24,28,29} (\sigma_3^{(l)})^{S_l}
\Gamma_* {\bf \Gamma}^{S_{32}}_5 {\bf C} \right)_{AB}
= \exp\{2\pi i K_{GSO} \} \left (\Gamma_* {\bf \Gamma}^
{\tilde{S}}_5 {\bf C} \right)_{AB},
\ee
where $\tilde{S}=S_{17}+S_{24}+S_{29}+S_{32}$ and
\bea
K_{GSO}  &=&   (k_{02}+k_{12}+k_{14}+k_{23}+k_{24}+k_{34}+ 1/2
)S_1 +\nonumber\\
& + & (k_{00}+k_{01}+k_{02}+k_{03}+k_{04}+k_{12}
+k_{23}+k_{24}+ 1/2 )S_{17} + \nonumber\\
&+& (k_{00}+k_{01}+k_{03}+k_{04}+k_{13}+k_{34}+1/2 )S_{24}
+\nonumber\\
&+& (k_{00}+k_{01}+k_{03}+k_{04}+k_{13}+k_{14}+1/2
)S_{29}.
\eea
At this point the amplitude is given by the following equation
\bea
T_{g=1} & = &
      \left( \frac{e^4}{\pi^6} \right)
\sum_{m_i,n_j} C^{\bbf {\alpha}}_{\bbf {\beta}} \ e^{2\pi i K_{GSO} }
{\bf W}_3^A {\bf V}_4^B \epsilon_1^{\mu}\epsilon_2^{\rho}
 \int {{\rm d}^2 \tau\over ({\rm Im}\tau)^2 } \
(\bar{\eta} (\bar{\tau}) )^{-24} (\eta(\tau))^{-12} \times \nonumber\\
     & & \int {\rm d}^2z_2 {\rm d}^2z_3 {\rm d}^2z_4 \
\frac{\sqrt{\omega(z_3)\omega(z_4)}}
{\bar{\omega} (\bar{z}_1) \omega(z_1)\omega (w)} \
exp\left[\sum_{i<j}(k_ik_j) G_B (z_i,z_j) \right]\nonumber\\
 & & \times{\cal T}_L \left[{}^{\bar{\alpha}}_{\bar{\beta}} \right]
(z_1,z_2,z_3,z_4,w)
\times {\cal T}_R \left[{}^{\alpha}_{\beta} \right]
(z_1,z_2,z_3,z_4,w)\ , \label{oneloop2}
\eea
where
\bea
{\cal T}_L & = &
\prod_{l=1-7,17} \bar{\Theta}
\left[{}^{\bar{\alpha}_{l}}_{\bar{\beta}_{l}} \right]
(\frac12 \bar{\nu}_{34} \vert \bar{\tau} )\times
\prod_{l=8}^{16} \prod_{l=18}^{22}\bar{\Theta} \left[
{}^{\bar{\alpha}_{l}}_{\bar{\beta}_{l}} \right] ( 0 \vert \bar{\tau} )
\times (\bar{E} (\bar{z}_3,\bar{z}_4))^{-2} \nonumber\\
                 & \times & \left\{ \partial_{\bar{z}_1}
\partial_{\bar{z}_2} \log \bar{E}(\bar{z}_1,\bar{z}_2)
+\frac14 \ \partial_{\bar{z}_1}\log \frac{\bar{E}(\bar{z}_1,\bar{z}_3)}
{\bar{E}(\bar{z}_1,\bar{z}_4)} \
 \partial_{\bar{z}_2}\log \frac{\bar{E}(\bar{z}_2,\bar{z}_3)}
{\bar{E}(\bar{z}_2,\bar{z}_4)} \right. \nonumber\\
                 &  & \quad+ \frac12 \frac{\bar{\omega}(\bar{z}_1)}{2\pi i} \
\partial_{\bar{z}_2}\log \frac{\bar{E}(\bar{z}_2,\bar{z}_3)}
{\bar{E}(\bar{z}_2,\bar{z}_4)}\partial_{\nu} \log \bar{\Theta}
\left[{}^{\bar{\alpha}_{17}}_{\bar{\beta}_{17}} \right]
(\nu\vert\bar{\tau})\vert_{\nu=\frac12\bar{\nu}_{34}} \nonumber\\
                 &  & \quad + \frac12
\frac{\bar{\omega}(\bar{z}_2)}{2\pi i} \
\partial_{\bar{z}_1}\log \frac{\bar{E}(\bar{z}_1,\bar{z}_3)}
{\bar{E}(\bar{z}_1,\bar{z}_4)} \partial_{\nu} \log \bar{\Theta}
\left[{}^{\bar{\alpha}_{17}}_{\bar{\beta}_{17}} \right]
(\nu\vert\bar{\tau})\vert_{\nu=\frac12\bar{\nu}_{34}} \nonumber\\
                 &  & \quad + \left. \frac{\bar{\omega}(\bar{z}_1)}{2\pi i}
\frac{\bar{\omega}(\bar{z}_2)}{2\pi i} \
\left(\bar{\Theta} \left[{}^{\bar{\alpha}_{17}}_{\bar{\beta}_{17}} \right]
(\frac{\scriptstyle 1}{\scriptstyle 2}\bar{\nu}_{34}\vert\bar{\tau})
\right)^{-1} \ \partial^2_{\nu}
\bar{\Theta}
(\nu\vert\bar{\tau})\vert_{\nu=\frac12\bar{\nu}_{34}} \right\},
\eea
and
\bea
{\cal T}_R & = &
\prod_{l=24,28,29,32}\Theta
\left[{}^{\alpha_{l}}_{\beta_{l}} \right] (
\frac{1}{2}\nu_{34} \vert \tau )
\prod_{l=23,25,26,27,30,31}\Theta
\left[{}^{\alpha_{l}}_{\beta_{l}} \right] (0 \vert \tau)
  \frac{(-1)^{S_{33}}}{\sqrt2 E(z_3,z_4)} \nonumber\\
&\times& \left\{ \left[ \sum_{j=1}^4 \sum_{n \neq 1} \sum_{p\neq 2}
 k_j^{\tau} k_n^{\mu} k_p^{\rho}
\partial_w G_B(z_j,w)\partial_1 G_B(z_n,z_1)\partial_2
G_B(z_p,z_2) \right. \right. \nonumber\\
& & \quad + g^{\mu \rho} \partial_1 \partial_2 G_B(z_1,z_2)
\sum_{j=1}^4 k_j^{\tau} \partial_w G_B(z_j,w)\nonumber\\
& & \quad + g^{\mu \tau} \partial_1 \partial_w G_B(z_1,w)
\sum_{j\neq 2} k_j^{\rho}\partial_2 G_B(z_j,z_2)\nonumber\\
& & \quad \left. +  g^{\rho \tau} \partial_2 \partial_w G_B(z_2,w)
\sum_{j \neq 1} k_j^{\mu}\partial_1 G_B(z_j,z_1)\right]
\left(\Gamma^{\tau} (\Gamma^5)^{\tilde{S}}
{\bf C}\right)_{AB}\nonumber\\
& - & k_1^{\nu} \left[ g^{\rho \tau} \partial_2 \partial_w G_B(z_2,w)
+ \sum_{i=1}^4 \sum_{j\neq 2}
 k_i^{\tau} k_j^{\rho}
\partial_w G_B(z_i,w) \partial_2 G_B(z_j,z_2)\right]\nonumber\\
& & \quad \times \left({\cal I}(w)\right)^{-1} {\cal I}(z_1)
\left[ G^-(z_1,w)
\left( \left(\Gamma^{\mu}\Gamma^{\nu}\Gamma^{\tau}
- g^{\mu \nu}\Gamma^{\tau} \right) (\Gamma^5)^{\tilde{S}}
{\bf C}\right)_{AB} \right. \nonumber\\
& & \quad + \left. \left(G^-(z_1,w) + G^+(z_1,w)\right)
\left( \left(g^{\mu\tau}\Gamma^\nu -g^{\nu\tau}\Gamma^\mu \right)
(\Gamma_{5})^{\tilde{S}}{\bf C}\right)_{AB} \right] \nonumber\\
& - & k_2^{\sigma} \left[ g^{\mu \tau} \partial_1 \partial_w G_B(z_1,w)
+ \sum_{i=1}^4 \sum_{j\neq 1}
 k_i^{\tau} k_j^{\mu}
\partial_w G_B(z_i,w) \partial_1 G_B(z_j,z_1)\right]\nonumber\\
& & \quad \times \left({\cal I}(w)\right)^{-1} {\cal I}(z_2)
\left[ G^-(z_2,w)
\left( \left(\Gamma^{\rho}\Gamma^{\sigma}\Gamma^{\tau}
- g^{\rho \sigma}\Gamma^{\tau} \right) (\Gamma^5)^{\tilde{S}}
{\bf C}\right)_{AB} \right. \nonumber\\
& & \quad + \left. \left(G^-(z_2,w) + G^+(z_2,w)\right)
\left( \left(g^{\rho \tau}\Gamma^\sigma -g^{\sigma\tau}\Gamma^\rho \right)
(\Gamma_{5})^{\tilde{S}}{\bf C}\right)_{AB} \right] \nonumber\\
    & - & k_1^{\nu} k_2^{\sigma}
\left[\sum_{j=1}^4 k_j^{\tau} \partial_w G_B(w,z_j)\right]
 \left[\left(G^+(z_1,z_2)\right)^2
\left((g^{\mu\rho}g^{\nu\sigma}-g^{\mu\sigma}g^{\nu\rho}) \Gamma^\tau
(\Gamma^{5})^{\tilde{S}}{\bf C}\right)_{AB}  \right. \nonumber\\
    &  & \quad + G^+(z_1,z_2)G^-(z_1,z_2)
\left( \left( g^{\nu\sigma}\Gamma^{\mu}\Gamma^{\rho}\Gamma^{\tau}
-g^{\nu\rho}\Gamma^{\mu}\Gamma^{\sigma}\Gamma^{\tau}\right.\right.\nonumber\\
& & \qquad +g^{\mu\rho}\Gamma^{\nu}\Gamma^{\sigma}\Gamma^{\tau}
-g^{\mu\sigma}\Gamma^{\nu}\Gamma^{\rho}\Gamma^{\tau}
-2(g^{\mu\rho}g^{\nu\sigma}-g^{\mu\sigma}g^{\nu\rho}) \Gamma^\tau
\nonumber\\
& & \qquad -(g^{\nu\sigma}g^{\rho\tau}-g^{\nu\rho}g^{\sigma\tau})\Gamma^\mu
+  (g^{\mu\sigma}g^{\rho\tau}-g^{\mu\rho}g^{\sigma\tau}) \Gamma^\nu
\nonumber\\
& & \qquad + \left. \left. (g^{\nu\sigma}g^{\mu\tau}-g^{\mu\sigma}g^{\nu\tau})
\Gamma^\rho -
(g^{\mu\tau}g^{\nu\rho}-g^{\mu\rho}g^{\nu\tau})
\Gamma^\sigma \right)
(\Gamma^{5})^{\tilde{S}}{\bf C}\right)_{AB}\nonumber\\
    & & \quad -\left({\cal I}(w)\right)^{-1}
{\cal I}(z_2)G^+(z_1,z_2)G^+(z_1,w)\nonumber\\
& & \qquad
\times\left(\left((g^{\nu\sigma}g^{\mu\tau}-g^{\mu\sigma}g^{\nu\tau})
\Gamma^\rho -
(g^{\mu\tau}g^{\nu\rho}-g^{\mu\rho}g^{\nu\tau})
\Gamma^\sigma \right)
(\Gamma^{5})^{\tilde{S}}{\bf C}\right)_{AB} \nonumber\\
    & & \quad -\left({\cal I}(w)\right)^{-1}
{\cal I}(z_2)G^-(z_1,z_2)G^+(z_1,w) \nonumber\\
& &\qquad \times \left( \left( g^{\mu\tau}\Gamma^{\nu}\Gamma^{\rho}
\Gamma^{\sigma}
-g^{\nu\tau}\Gamma^{\mu}\Gamma^{\rho}\Gamma^{\sigma}
+ g^{\nu\tau}g^{\rho\sigma}\Gamma^{\mu}
-g^{\mu\tau}g^{\rho\sigma}\Gamma^{\nu} \right. \right.\nonumber\\
& & \qquad + \left. \left.(g^{\nu\sigma}g^{\mu\tau}-g^{\mu\sigma}g^{\nu\tau})
\Gamma^\rho
- (g^{\mu\tau}g^{\nu\rho}-g^{\mu\rho}g^{\nu\tau}) \Gamma^\sigma \right)
(\Gamma^{5})^{\tilde{S}}{\bf C}\right)_{AB} \nonumber\\
      & & \quad  + \left({\cal I}(w)\right)^{-1}
{\cal I}(z_1)G^+(z_1,z_2)G^+(z_2,w)\nonumber\\
& & \qquad \times\left(\left( (g^{\nu\sigma}g^{\rho\tau}-g^{\nu\rho}
g^{\sigma\tau}) \Gamma^\mu -
(g^{\mu\sigma}g^{\rho\tau}-g^{\mu\rho}g^{\sigma\tau}) \Gamma^\nu \right)
(\Gamma^{5})^{\tilde{S}}{\bf C}\right)_{AB} \nonumber\\
      &  & \quad - \left({\cal I}(w)\right)^{-1}
{\cal I}(z_1)G^-(z_1,z_2)G^+(z_2,w)\nonumber\\
& &\qquad \times \left( \left( g^{\rho\tau}\Gamma^{\mu}\Gamma^{\nu}
\Gamma^{\sigma}
-g^{\sigma\tau}\Gamma^{\mu}\Gamma^{\nu}\Gamma^{\rho}
+ g^{\mu\nu}g^{\sigma\tau}\Gamma^{\rho}
-g^{\mu\nu}g^{\rho\tau}\Gamma^{\sigma}\right.\right. \nonumber\\
& & \qquad  -\left. \left. \left. \left.
(g^{\nu\sigma}g^{\rho\tau}-g^{\nu\rho}g^{\sigma\tau}) \Gamma^\mu +
(g^{\mu\sigma}g^{\rho\tau}-g^{\mu\rho}g^{\sigma\tau}) \Gamma^\nu \right)
(\Gamma^{5})^{\tilde{S}}{\bf C} \right)_{AB}\right]\right\}
\eea
As already mentioned in Section 3, BRST-invariance of the vertex operators
describing physical states implies on-shell conditions for the external
states, as well as the transversality condition for the photons and
the Dirac equation for the fermions:
\bea
k_1^2 = &k_2^2 &=k_3^2=k_4^2=0,\nonumber\\
\epsilon \cdot k &= &0, \nonumber\\
{\bf W}_3^T \ds{k_3} & = &\ds{k_4}{\bf C}{\bf V}_4=0 \label{onshell}.
\eea
These constraints were used to derive the explicit form of the
vertex operators (\ref{photon}) and  (\ref{fermion}),
but we were careful not to explicitly use them
in the computation leading to eq.\ (\ref{oneloop2}).
Apart from technical advantages in doing the computations,
the reasons for this choice are that this expression for the amplitude
is somewhat more compact than the final one we will show in the next
section and, being somehow ``off-shell'', it could turn out to be
useful as a starting point for the analysis of the field theory
limit of the scattering amplitude, where one usually faces the
problem of ``going off-shell'' (see for instance \cite{Kap,BK,DiV}).
Of course, not being on-shell, the amplitude (\ref{oneloop2}) is not gauge
nor conformal invariant or independent on the position of insertion
of the PCO operator.
\section{The Final Form of the Amplitude}
\setcounter{equation}{0}
The momentum conservation
$k_1+k_2+k_3+k_4=0$ and on-shell conditions (\ref{onshell})
allow us to considerably simplify the Lorentz structures
of the amplitude (\ref{oneloop2}).
Expanding all the products, several terms in expression
(\ref{oneloop2}) drop out because of the on-shell and transversality
conditions.
Then we eliminate $k_2$ using momentum conservation, and we rearrange all
the products of gamma matrices contracted with the momenta $k_3$ and
$k_4$ in such a way to have the factor $\ds{k_3}$ ($\ds{k_4}$) always
on the left (right).
At this point all terms involving $\ds{k_3}$ and $\ds{k_4}$ vanish
because of Dirac equations, and we obtain the
final form of the on-shell amplitude:
\bea
T_{g=1} & &(\ e^{\pm} \ + \ \gamma  \ \rightarrow \ e^{\pm} \
+ \ \gamma )= \nonumber\\
     & & \left( \frac{e^4}{\pi^6} \right)
\sum_{m_i,n_j} C^{\bbf {\alpha}}_{\bbf {\beta}} \ e^{2\pi i
K_{GSO} }
\int {{\rm d}^2 \tau\over ({\rm Im}\tau)^2 } \
(\bar{\eta} (\bar{\tau}) )^{-24} (\eta(\tau))^{-12} \times
\nonumber\\
     & & \int {\rm d}^2z_2 {\rm d}^2z_3 {\rm d}^2z_4 \
\frac{\sqrt{\omega(z_3)\omega(z_4)}}
{\bar{\omega} (\bar{z}_1) \omega(z_1)\omega (w)} \
exp\left[\sum_{i<j}(k_ik_j) G_B (z_i,\bar{z_i},z_j,\bar{z_j})
\right] \nonumber\\
 & & \times{\cal T}_L
\left[{}^{\bar{\bbf{\alpha}}}_{\bar{\bbf{\beta}}} \right]
(z_1,z_2,z_3,z_4,w)
\times {\cal T}_R \left[{}^{\bbf{\alpha}}_{\bbf{\beta}} \right]
(z_1,z_2,z_3,z_4,w). \label{result}
\eea
where
\bea
{\cal T}_L & (& z_1,z_2,z_3,z_4,w) =
\prod_{l=1-7,17} \bar{\Theta}
\left[{}^{\bar{\alpha}_{l}}_{\bar{\beta}_{l}} \right]
(\frac12 \bar{\nu}_{34} \vert \bar{\tau} )
\prod_{l=8}^{16} \prod_{l=18}^{22}\bar{\Theta} \left[
{}^{\bar{\alpha}_{l}}_{\bar{\beta}_{l}} \right] ( 0 \vert
\bar{\tau} )
\times (\bar{E} (\bar{z}_1,\bar{z}_2))^{-2} \nonumber\\
                 & \times & \left\{ \partial_{\bar{z}_1}
\partial_{\bar{z}_2} \log \bar{E}(\bar{z}_1,\bar{z}_2)
+\frac14 \ \partial_{\bar{z}_1}\log
\frac{\bar{E}(\bar{z}_1,\bar{z}_3)}
{\bar{E}(\bar{z}_1,\bar{z}_4)} \
 \partial_{\bar{z}_2}\log \frac{\bar{E}(\bar{z}_2,\bar{z}_3)}
{\bar{E}(\bar{z}_2,\bar{z}_4)} \right. \nonumber\\
                 &  & \quad+ \frac12
\frac{\bar{\omega}(\bar{z}_1)}{2\pi i} \
\partial_{\bar{z}_2}\log \frac{\bar{E}(\bar{z}_2,\bar{z}_3)}
{\bar{E}(\bar{z}_2,\bar{z}_4)}\partial_{\nu} \log \bar{\Theta}
\left[{}^{\bar{\alpha}_{17}}_{\bar{\beta}_{17}} \right]
(\nu\vert\bar{\tau})\vert_{\nu=\frac12\bar{\nu}_{34}}. \nonumber\\
                 &  & \quad + \frac12
\frac{\bar{\omega}(\bar{z}_2)}{2\pi i} \
\partial_{\bar{z}_1}\log \frac{\bar{E}(\bar{z}_1,\bar{z}_3)}
{\bar{E}(\bar{z}_1,\bar{z}_4)} \partial_{\nu} \log \bar{\Theta}
\left[{}^{\bar{\alpha}_{17}}_{\bar{\beta}_{17}} \right]
(\nu\vert\bar{\tau})\vert_{\nu=\frac12\bar{\nu}_{34}} \nonumber\\
                 &  & \quad + \left.
\frac{\bar{\omega}(\bar{z}_1)}{2\pi i}
\frac{\bar{\omega}(\bar{z}_2)}{2\pi i} \
\left(\bar{\Theta} \left[{}^{\bar{\alpha}_{17}}_{\bar{\beta}_{17}}
\right]
(\frac12\bar{\nu}_{34}\vert\bar{\tau}) \right)^{-1} \
\partial^2_{\nu}
\bar{\Theta}
(\nu\vert\bar{\tau})\vert_{\nu=\frac12\bar{\nu}_{34}} \right\},
\label{LmP2}
\eea
\bea
{\cal T}_R  & (  & z_1,z_2,z_3,z_4,w) =
\prod_{l=24,28,29,32}\Theta
\left[{}^{\alpha_{l}}_{\beta_{l}} \right] (\frac12 \nu_{34} \vert
\tau )
\prod_{l=23,25,26,27,30,31}\Theta
\left[{}^{\alpha_{l}}_{\beta_{l}} \right] (0 \vert
\tau)\nonumber\\
                 &\times&  \frac{(-1)^{S_{33}}}{\sqrt2}
(E(z_3,z_4))^{-1}
\left\{{\bf W}^T_3 \ds{k_1}\ds{\epsilon_1}\ds{\epsilon_2}
\left( \Gamma^5 \right)^{\tilde{S}}
{\bf C V}_4  \ {\cal A}_1(z_1,z_2,z_3,z_4,w) \right. +\nonumber\\
                 &  & \quad + {\bf W}^T_3 \ds{k_1}
\left( \Gamma^5 \right)^{\tilde{S}}
{\bf C V}_4  \ {\cal A}_2(z_1,z_2,z_3,z_4,w) + \nonumber\\
                 &  & \quad+ {\bf W}^T_3  \ds{\epsilon_1}
\left( \Gamma^5 \right)^{\tilde{S}}
{\bf C V}_4 \ {\cal A}_3(z_1,z_2,z_3,z_4,w) + \nonumber\\
                 &  & \quad+ \left.  {\bf W}^T_3 \ds{\epsilon_2}
\left( \Gamma^5 \right)^{\tilde{S}}
{\bf C V}_4 \ {\cal A}_4(z_1,z_2,z_3,z_4,w) \right\}\ .
\label{RmP2}
\eea
The factors ${\cal A}_i$, multiplying the gamma matrices, are
functions of the world-sheet coordinates, kinematics invariants and
polarization vectors. They are given explicitly in Appendix C.

Notice that, except for the factors
$exp\left[\sum_{i<j}(k_ik_j) G_B (z_i,\bar{z_i},z_j,\bar{z_j})
\right]$,
all the dependence on the external momenta, spinors and
polarizations is in the function ${\cal T}_R$.
As expected, after all Lorentz algebra has been done, only
four independent gamma matrix structures remain.

The amplitude (\ref{result}) is formally hermitian and presents divergences
in the integration over the moduli \cite{DP2,Ama}. It becomes absolutely
convergent only for purely
imaginary values of the Mandelstam variables $s$,$t$ and $u$
\cite{DP2,Mont,Gross}. These divergences are typical of string amplitudes
in the Lorentz covariant formulation. The physical interpretation of
such divergences has been discussed for instance in refs.\
\cite{PR2},\cite{PR5},\cite{PRlon},\cite{DP2},\cite{Mont},\cite{Wein}---\cite{Bere},
and is related to the unitarity of the scattering amplitudes.
Indeed the two problems of these amplitudes, divergences and formal
hermiticity instead of unitarity, are strictly related and can be
cured at the same time. One can regularize the integrals by an
analytical continuation in the external momenta so that at the same
time the correct poles and branch
cuts required by unitarity appear. In refs. \cite{DP2,Mont,Bere} one
can find examples of such analytical continuation procedure (the
stringy version of the Feynman $+i\epsilon$).\par

The amplitude presents also infrared divergences due to the presence
of massless states. As in field theory, these divergences correspond
to the emission/absorption of soft photons/electrons, and can be
removed by introducing an infrared cut-off.\par

On the other hand, as any string amplitude, it is free from ultraviolet
divergences and automatically supplies a regulator for the chiral massless
fermions. Notice however that these divergences will reappear in the
field theory limit as divergences in $\alpha'\rightarrow0$.\par
The dependence of our result on the PCO variable $w$ deserves some comments.
It is well known that the amplitude should not depend on a PCO
insertion point $w$. Therefore its derivative with respect to  $w$
must be zero. In general this comes about only
after the integration over the moduli space has been
performed, since the differentiation with respect to $w$ gives rise to
a total derivative in the integrand.
However, in this case, it is possible to show that the integrand itself
is independent on the PCO's insertion point.
Consider the amplitude (\ref{oneloop1}), where we absorbed two PCO's in
the photon vertices, and
write the remaining PCO as $\Pi(w)=2\{Q_{BRST},\xi(w)\}$. Then moving the BRST
commutator onto the other operators, we get
\bea
\partial_w T^{1-loop} &= & 2 C_{g=1}
      \int {\rm d}^2\tau {\rm d}^2z_2 {\rm d}^2z_3
      {\rm d}^2z_4 \
      \sum_{m_i,n_j} C^{\bbf {\alpha}}_{\bbf {\beta}} \
      \sum_{m^I=\tau,z_2,z_3,z_4} \frac{\partial}{\partial m^I}\nonumber\\
  & &\wew{(\bar{\eta}_{\bar{\tau}}\vert \bar{b})
         (\bar{\eta}_{\bar{z}_2}\vert \bar{b})
         (\bar{\eta}_{\bar{z}_3}\vert \bar{b})
         (\bar{\eta}_{\bar{z}_4}\vert \bar{b})
         \frac{\partial}{\partial(\eta_{m^I})}
         \{(\eta_{z_2}\vert b)
         (\eta_{z_3}\vert b)(\eta_{z_4}\vert b)\} \nonumber\\
  &\times&  \vert c(z_1)c(z_2)c(z_3)c(z_4)\vert^2 \
      \partial_w \xi(w) \ {\cal V}^{(0)}_{\rm photon}
      (z_1,\bar{z}_1;k_1;\epsilon_1)
      {\cal V}^{(0)}_{\rm photon}
(z_2,\bar{z}_2;k_2;\epsilon_2)\nonumber\\
  &\times & {\cal V}^{(-1/2)} (z_3,\bar{z}_3;k_3;{\bf
W}_3) \
      {\cal V}^{(-1/2)} (z_4,\bar{z}_4;k_4;{\bf V}_4)},
\label{derivative}
\eea
where we used the fact that $\bar{c}c{\cal V}$ is BRST invariant
and that \cite{Mart}
\be
\wew{\{Q_{BRST},(\eta_I\vert b)\} \ldots} =
\wew{(\eta_I\vert T_B) \ldots} =  \frac{\partial}{\partial m^I}\wew{\ldots}.
\ee
Then superghost charge conservation forces the derivative
(\ref{derivative}) to vanish altogether,
meaning that the integrand must be explicitly independent of $w$.\par
As a check of the correctness of our computation, we can then verify
that the integrand eq.(\ref{RmP2}) is indeed independent of $w$.
We could compute its derivative with respect
to $w$ and see if it gives zero.
However, in the case at hand, it is not so straightforward to prove
this statement explicitly due to the identities in theta-functions
we would need to prove.
Instead we proved that the quantity
${1\over \omega(w)}{\cal T}_{R}(w)$, which is the only factor in the
amplitude depending on $w$, is a meromorphic function
of $w$ on the torus, that it does not have zeros and that the
residues at poles
vanish. Thus it is a constant (as a function of $w$)
and hence independent of $w$.\par

Finally we can try to compare the string
scattering amplitude with the corresponding one in field theory.
The amplitude (\ref{result}) corresponds
to a field theory amplitude where the integral over loop
momenta and the Lorentz algebra have been already done.
The integrals over the moduli correspond to
the integrals over the Schwinger proper-times.
At first glance, one notices that the string amplitude contains more
kinematical structures than the field theory one
(see for example \cite{Fey}). However a precise comparison between
them would require the explicit evaluation of the integrals over
the Schwinger parameters in field theory and the integrals over
moduli as well as  the sum over spin-structures in string theory.
While in field theory the integrals can be done, the same is usually
not true in string theory. This prevents us from making any particular
claim about the properties of the string-theory before making any
suitable approximation or taking the field theory limit.
%
%
%
%
%
\section*
{Acknowledgements}
It is a pleasure to thank Kaj Roland and Luciano Girardello for useful
discussions.\par
This work is partially supported by the European Commission TMR program
ERBFMRX-CT96-0045 in which A.P.\ is associated to the Institute for
Theoretical Physics, K.U.\ Leuven, and M.P.\ is associated to the
Physics Department, Milano University.
A.P.\ would like to thank CERN for its hospitality while part of this
work was carried out.
\section*
{Appendices}
\appendix

\section {Notations, Conventions and Useful Formulae}
\setcounter{equation}{0}
In this appendix we will give our conventions for the operator fields,
partition functions and correlation functions on the torus.
First of all we state our conventions for
the Dedekind $\eta$-function, the
theta functions and the prime form.
The Dedekind $\eta$-function is given by
\be
\eta(\tau) = k^{1/24} \prod_{n=1}^{\infty} (1-k^n),  \qquad
k = e^{2\pi i \tau},
\label{adedekind}
\ee
and our conventions for the theta functions are
\bea
\Teta{\alpha}{\beta}(\nu\vert\tau) \, & = & e^{i\pi(\frac12-\alpha)^2\tau}
e^{2\pi i(\frac12 +\beta)(\frac12-\alpha)} e^{2\pi i(\frac12
-\alpha)\nu}\ \times  \label{atheta} \\
& &  \prod_{n=1}^\infty
(1-k^n)(1-k^{n+\alpha-1}e^{-2\pi i (\beta+\nu)}) (1-k^{n-\alpha}
e^{2\pi i (\beta+\nu)}) \nonumber\\
& = &  \sum_{r\in{\Bbb Z}} e^{\pi i (r+\frac12-\alpha)^2\tau
+2\pi i (r+\frac12-\alpha)(\nu +\beta+\frac12)} \nonumber\\
\Theta_1\equiv\Teta00,& &\quad \Theta_2\equiv \Teta{\ 0}{1/2}\ ,\quad
\Theta_3\equiv\Teta{1/2}{1/2} \ ,\quad
\Theta_4 \equiv \Teta{1/2}{\ 0}. \nonumber\\
\eea
The standard Riemann identity is
\be
\sum_{\alpha,\beta} e^{2\pi i(\alpha+\beta)} \prod_{i=1}^4
\Teta\alpha\beta(x_i\vert\tau) \ =\ 0,
\label{acfor}
\ee
where  $\alpha,\beta\, = \{0,\frac12\}$ and
one of the following equations must hold
\bea
x_1+x_2+x_3+x_4 &=& 0 \qquad\qquad\quad
x_1-x_2-x_3+x_4=0  \nonumber\\
x_1-x_2+x_3-x_4 & = & 0 \qquad\qquad\quad x_1+x_2-x_3-x_4 = 0.
\eea
The prime form is
\be
E(z,w) = {2\pi i \Theta_1(\nu_{zw}
\vert \tau) \over \sqrt{\omega(z)\omega(w)}
\Theta^\prime_1 (0\vert \tau) } \, , \qquad \nu_{zw} =
\int_{w}^{z} {\omega \over 2\pi i } \,
\label{aprimef}
\ee
where $\omega(z)$ is the holomorphic 1-form on the torus, normalized to
have period $2\pi i$ around the $a$-cycle. In the parametrization where
$\omega(z)=1/z$ the prime form (\ref{aprimef}) becomes
\be
E(z,w) = (z-w) \prod_{n=1}^\infty {(1-{z\over w} k^n )
(1-{w\over z}k^n) \over (1-k^n)^2 }.
\ee

The space-time coordinate fields $X^{\mu}$ have mode expansion
\be
X^{\mu}(z,\bar{z}) \,= \, q^{\mu} -ik^{\mu}\log(z,\bar{z}) +
i \sum_{n\neq0}\frac{\alpha_n^{\mu}}{n}z^{-n}
+ i \sum_{n\neq0}\frac{\bar{\alpha}_n^{\mu}}{n}\bar{z}^{-n}
\ee
and satisfy the OPE
\be
X^\mu (z,\bar{z}) X^\nu (w,\bar{w}) \eqope - g^{\mu\nu}
\left(\log(z-w) + \log (\bar{z}-\bar{w}) \right)
+ \ldots \ .
\ee

Their one-loop partition function is given by
\be
Z_{X} = \prod_{n=1}^{\infty} \vert 1 - k^n \vert^{-8} (2\pi {\rm Im}
\tau)^{-2},
\ee
and the genus one correlator is
\be
\wew{X^\mu(z,\bar{z}) X^\nu(w,\bar{w})} \, =\, - g^{\mu\nu} \, G_B
(z,\bar{z};w,\bar{w}) \ Z_{X}.
\ee
Here $G_B$ is the bosonic Green function on the torus where the non-zero
mode part of the partition function has been removed:
\be
G_B(z,\bar{z};w,\bar{w}) =
2\left[ \log \vert
E(z,w)\vert - \frac12 {\rm Re}\left( \int_{w}^ {z}\omega
\right)^2
{1\over 2\pi {\rm Im}\tau} \right]
\label{bgf}
\ee
The real world-sheet fermions associated with the space-time
coordinates (eq.(~\ref{stf})) have mode expansion
\be
\psi^{\mu}(z) \,= \,
\sum_{n}\psi_n^{\mu}z^{-n-1/2}, \qquad\quad
\{\psi^{\mu},\psi^{\nu}\}=g^{\mu\nu} \delta_{m+n,0}
\ee
with $n$ integer (half-integer) for Ramond (Neveu-Schwarz)
boundary conditions. Their OPE is
\be
\psi^\mu(z) \psi^\nu(w) \eqope
\frac{g^{\mu \nu}} {z-w} + \ldots \,.
\ee
The real internal world-sheet fermions (eq.(~\ref{intf})) have a similar mode
expansion and their OPE is given by
\be
\psi^m_{(l)}(z) \psi^n_{(k)}(w) \eqope
\frac{\delta^{m,n}\delta_{l,k}}{z-w} \, + \, \ldots.
\ee
Correlations functions are defined according to
eq.(~\ref{gencorr}) and are
computed bosonizing all complex fermions according to eq.(\ref{bosonization}).
The fundamental genus one correlator~\cite{PDV1} is
\be
\vev{\prod_{i=1}^N e^{q_i \phi(z_i)} }
\left[{}^\alpha_\beta\right] =
\delta_{\sum_{i=1}^N q_i,0} \prod_{i<j} \left[ E(z_i,z_j)
\right]^{q_iq_j} \, \Teta\alpha\beta \left(\sum_{i=1}^N q_i \int^{z_i}
\frac\omega{2\pi i}
\vert \tau \right),  \label{aboscorr}
\ee
where we have
explicitly displayed the spin structure dependence of the
correlator, whereas in the paper we often adopt the following shorthand
notation
\be
\vev{\prod_{i=1}^N e^{q_i \phi(z_i)} }_{(l)}\ =\
\vev{\prod_{i=1}^N e^{q_i \phi(z_i)} } \left[{}^{\alpha_l}_{\beta_l}
\right].
\ee
For the reparametrization ghosts we follow the conventions of
ref.~\cite{FMS}.
The normalization of the partition function is the standard one and
the explicit expression can be found in refs.~\cite{Kaj2,PDV1}.
The correlator relevant for our $1$-loop scattering amplitude
involving $N$ physical external states is
\bea
{\rm d}^2 k & & \prod_{i=1}^{N-1} {\rm d}^2 z_i \
\wew{ \left| (\eta_k \vert b) \prod_{i=1}^{N-1}
(\eta_{z_i} \vert b) \prod_{i=1}^N c(z_i) \right|^2} \label{bone} \\
& & \qquad  = {{\rm d}^2 k \over \bar{k}^2 k^2} \prod_{i=1}^{N-1}
{\rm d}^2 z_i \left| {1 \over \omega(z_N)} \right|^2 \prod_{n=1}^{\infty}
\left| 1 - k^n \right|^4.
\eea
For the superghosts mode expansions, OPE and bosonization we follow
the standard conventions of ref.~\cite{FMS} (see also eqq.
(\ref{modeexp}), (\ref{bossupgh})).
We always remain inside the little algebra; whit our definition of the
theta functions, the partition function for the superghost
is~\cite{PR1,PR2}
\bea
\wew{\prod_{i=1}^N e^{q_i
\phi(z_i)} } \ & = & (-1)^S k^{1/2}
\prod_{n=1}^{\infty} (1-k^n) \prod_{i=1}^N (\omega(z_i))^{-q_i} \times
\nonumber\\
& & \prod_{i < j} (E(z_i,z_j))^{-q_i q_j} \,
\left[ \Teta{\alpha}{\beta} \left( \sum_{j=1}^N q_j
\int_{z_0}^{z_j}
{\omega \over 2\pi i } \vert \tau \right) \right]^{-1}.
\label{bseven}
\eea
\section {Theta Functions Identities}
\setcounter{equation}{0}
In this section we give the list of the identities in the Theta
functions which arise in computing the correlators of Section 3.
\be
{\cal A}\left[{}^{\alpha}_{\beta} \right](z_1,z_2,w)
\Ical{\alpha}{\beta}{(w)} =
\GGp{\alpha}{\beta}{(z_1,z_2)} \,
{\cal D}_+\left[{}^{\alpha}_{\beta} \right](z_1,z_2,w),\label{identities}
\ee
$$
{\cal B}\left[{}^{\alpha}_{\beta} \right](z_1,z_2)
\Ical{\alpha}{\beta}{(w)} =
\GGp{\alpha}{\beta}{(z_1,z_2)} \,
{\cal E}_2 \left[{}^{\alpha}_{\beta} \right](z_1,z_2,w),
$$
$$
{\cal C}_-\left[{}^{\alpha}_{\beta} \right](z_1,z_2,w) =
- \GGp{\alpha}{\beta}{(z_1,z_2)}\,
\GGp{\alpha}{\beta}{(z_1,w)},
$$
$$
{\cal C}_+\left[{}^{\alpha}_{\beta} \right](z_1,z_2,w)
\Ical{\alpha}{\beta}{(z_2)} =
\GGm{\alpha}{\beta}{(z_1,z_2)} \,
{\cal D}_+\left[{}^{\alpha}_{\beta} \right](z_1,z_2,w),
$$
$$
{\cal D}_-\left[{}^{\alpha}_{\beta} \right](z_1,z_2,w) =
\Ical{\alpha}{\beta}{(z_2)}
\GGp{\alpha}{\beta}{(z_1,w)},
$$
$$
{\cal D}_+\left[{}^{\alpha}_{\beta} \right](z_1,z_2,w) =
{\cal E}_1\left[{}^{\alpha}_{\beta} \right](z_1,z_2,w)
- {\cal D}_-\left[{}^{\alpha}_{\beta} \right](z_1,z_2,w) +
{\cal E}_2\left[{}^{\alpha}_{\beta} \right](z_1,z_2,w)
$$
where the functions $G^{\pm}$ and ${\cal I}$ are defined in Section 3,
eqq.(\ref{eIIGG}), and \footnotemark
\footnotetext{~Even if not explicitly written, all functions listed
below and in the following Appendix depend also on the world-sheet
coordinates $z_3,z_4$. For definitions of theta functions and prime
form see Appendix A.}
\bea
&{\cal A} & \left[{}^{\alpha}_{\beta} \right](z_1,z_2,w) =
\left\{ \partial_{z_1}
\partial_{z_2}  \log E(z_1,z_2)
+ \frac{\omega(z_1)}{2\pi i}
\frac{\omega(z_2)}{2\pi i} \ \partial^2_{\nu} \log
\Theta (\nu\vert\tau)\vert_{\nu=\mu_w} \right. \label{definition2}\\
   & & + \left(\partial_{z_1}
\log \frac{E(z_1,w)}{\sqrt{E(z_1,z_3)E(z_1,z_4)}} \
+ \frac{\omega(z_1)}{2\pi i}
\partial_{\nu} \log \Theta
\left[{}^{\alpha}_{\beta} \right]
(\nu\vert\tau)\vert_{\nu=\mu_w} \right) \nonumber\\
& & \times  \left. \left(\partial_{z_2}
\log \frac{E(z_2,w)}{\sqrt{E(z_2,z_3)E(z_2,z_4)}} \
+ \frac{\omega(z_2)}{2\pi i}
\partial_{\nu} \log \Theta
\left[{}^{\alpha}_{\beta} \right]
(\nu\vert\tau)\vert_{\nu=\mu_w} \right) \right\},\nonumber
\eea
\bean
&{\cal B} & \left[{}^{\alpha}_{\beta} \right](z_1,z_2) =
\left\{ \partial_{z_1}
\partial_{z_2}  \log E(z_1,z_2)
+\frac14 \ \partial_{z_1} \log \frac{E(z_1,z_3)}{E(z_1,z_4)} \
 \partial_{z_2}\log \frac{E(z_2,z_3)}{E(z_2,z_4)} \right. \\
                 &  & + \frac12 \frac{\omega(z_1)}{2\pi i} \
\partial_{z_2}\log \frac{E(z_2,z_3)}
{E(z_2,z_4)}\partial_{\nu} \log \Theta
\left[{}^{\alpha}_{\beta} \right]
(\nu\vert\tau)\vert_{\nu=\frac12\nu_{34}} \\
                 &  & + \frac12
\frac{\omega(z_2)}{2\pi i} \
\partial_{z_1}\log \frac{E(z_1,z_3)}
{E(z_1,z_4)} \partial_{\nu} \log \Theta
\left[{}^{\alpha}_{\beta} \right]
(\nu\vert\tau)\vert_{\nu=\frac12\nu_{34}} \\
                 &  & + \left. \frac{\omega(z_1)}{2\pi i}
\frac{\omega(z_2)}{2\pi i} \
\left(\Theta \left[{}^{\alpha}_{\beta} \right]
(\frac12\nu_{34}\vert\tau) \right)^{-1} \ \partial^2_{\nu}
\Theta (\nu\vert\tau)\vert_{\nu=\frac12\nu_{34}} \right\},
\eean
\bean
&{\cal C}_{\pm}& \left[{}^{\alpha}_{\beta} \right](z_1,z_2,w) =
\frac12 \left(E(z_2,w)\right)^{-1}
\left\{ \sqrt{\frac{E(z_3,w) E(z_2,z_4)}
{E(z_2,z_3)E(z_4,w)}}
\frac{\Theta \left[{}^{\alpha}_{\beta} \right] (\rho_{w,z_2}\vert \tau)}
{\Theta \left[ {}^{\alpha}_{\beta} \right]
(\frac12 \nu_{34} \vert\tau)} \right. \\
& & \times
\left[\partial_{z_1} \log \left(\frac{E(z_1,w)}{E(z_1,z_2)}
\sqrt{\frac{E(z_1,z_3)}{E(z_1,z_4)}}\right)
+ \frac{\omega(z_1)}{2\pi i}
\partial_{\nu} \log \Theta
\left[{}^{\alpha}_{\beta} \right]
(\nu\vert\tau)\vert_{\nu=\rho_{w,z_2}} \right] \\
& & \pm  \sqrt{\frac{E(z_2,z_3)E(z_4,w)}
{E(z_3,w)E(z_2,z_4)}}
\frac{\Theta \left[{}^{\alpha}_{\beta} \right] (\rho_{z_2,w}\vert \tau)}
{\Theta \left[ {}^{\alpha}_{\beta} \right]
(\frac12 \nu_{34} \vert\tau)} \\
& & \times  \left.
\left[\partial_{z_1} \log \left(\frac{E(z_1,z_2)}{E(z_1,w)}
\sqrt{\frac{E(z_1,z_3)} {E(z_1,z_4)}}\right)
+ \frac{\omega(z_1)}{2\pi i}
\partial_{\nu} \log \Theta
\left[{}^{\alpha}_{\beta} \right]
(\nu\vert\tau)\vert_{\nu=\rho_{z_2,w}} \right] \right\},
\eean
\bean
&{\cal D}_{\pm}& \left[{}^{\alpha}_{\beta} \right](z_1,z_2,w) =
\frac12 \sqrt{E(z_3,z_4)}
\left({\Theta \left[{}^{\alpha}_{\beta} \right]
(\frac12\nu_{34}\vert\tau)}\right)^{-1} \\
& & \left\{ \sqrt{\frac{E(z_2,z_3) E(z_2,z_4)}
{E(z_1,z_3) E(z_1,z_4)E(w,z_3) E(w,z_4)}}
\frac{E(z_1,w)}{E(z_2,z_1)E(z_2,w)}
{\Theta \left[{}^{\alpha}_{\beta} \right]
(\mu_{\frac{z_1w}{z_2}}\vert\tau)} \right.\\
& & \pm
\sqrt{\frac{E(z_1,z_3) E(z_1,z_4)}
{E(z_2,z_3) E(z_2,z_4)E(w,z_3) E(w,z_4)}}
\frac{E(z_2,w)}{E(z_1,z_2)E(z_1,w)}
{\Theta \left[{}^{\alpha}_{\beta} \right]
(\mu_{\frac{z_2w}{z_1}}\vert\tau)} \\
& & \pm
\left. \sqrt{\frac{E(w,z_3) E(w,z_4)}
{E(z_1,z_3) E(z_1,z_4)E(z_2,z_3) E(z_2,z_4)}}
\frac{E(z_1,z_2)}{E(w,z_1)E(w,z_2)}
{\Theta \left[{}^{\alpha}_{\beta} \right]
(\mu_{\frac{z_1z_2}{w}}\vert\tau)} \right],
\eean
\bean
&{\cal E}_{1,2}& \left[{}^{\alpha}_{\beta} \right](z_1,z_2,w) =
\frac12 \sqrt{E(z_3,z_4)}
\left({\Theta \left[{}^{\alpha}_{\beta} \right]
(\frac12\nu_{34}\vert\tau)}\right)^{-1} \\
& & \left\{ \sqrt{\frac{E(z_2,z_3) E(z_2,z_4)}
{E(z_1,z_3) E(z_1,z_4)E(w,z_3) E(w,z_4)}}
\frac{E(z_1,w)}{E(z_2,z_1)E(z_2,w)}
{\Theta \left[{}^{\alpha}_{\beta} \right]
(\mu_{\frac{z_1w}{z_2}}\vert\tau)} \right.\\
& & \mp
\sqrt{\frac{E(z_1,z_3) E(z_1,z_4)}
{E(z_2,z_3) E(z_2,z_4)E(w,z_3) E(w,z_4)}}
\frac{E(z_2,w)}{E(z_1,z_2)E(z_1,w)}
{\Theta \left[{}^{\alpha}_{\beta} \right]
(\mu_{\frac{z_2w}{z_1}}\vert\tau)} \\
& & \pm
\left. \sqrt{\frac{E(w,z_3) E(w,z_4)}
{E(z_1,z_3) E(z_1,z_4)E(z_2,z_3) E(z_2,z_4)}}
\frac{E(z_1,z_2)}{E(w,z_1)E(w,z_2)}
{\Theta \left[{}^{\alpha}_{\beta} \right]
(\mu_{\frac{z_1z_2}{w}}\vert\tau)} \right],
\eean
\section {Lorentz Functions of the On-Shell Amplitude}
\setcounter{equation}{0}
In this appendix we display the explicit expressions for the
coefficients ${\cal A}_i$ which appear in the amplitude ${\cal T}_R$
of eq.(\ref{RmP2})
They are functions of the external
momenta,
polarization vectors and world sheet coordinates.\footnotemark
\footnotetext{All functions listed
here depend also on the coordinates $z_3,z_4$.}
\bea
{\cal A}_1(&z_1&,z_2,w) = (k_1k_2) \left[\partial_w G_B(w,z_1)
- \partial_w G_B(w,z_2)\right]G^+(z_1,z_2)G^-(z_1,z_2) \label{definition3}\\
                     & + &  \frac12 \sum_{j=1}^4 (k_1k_j)
\partial_{z_1} G_B(z_j,z_1) \partial_w G_B(z_1,w) I(z_2) + \nonumber\\
                     & + & \frac12 \sum_{j=1}^4 (k_2k_j)
\partial_{z_2}G_B(z_j,z_2)\partial_w G_B(z_2,w)I(z_1)  + \nonumber\\
                     & + & \sum_{j=1}^4 \partial_w G_B(w,z_j)
\left[(k_1k_j){\cal I}(z_2)G^+(z_1,w)
+(k_2k_j){\cal I}(z_1)G^+(z_2,w)\right]
\frac{G^-(z_1,z_2)}{{\cal I}(w)}, \nonumber
\eea
\bean
{\cal A}_2 ( & z_1, & z_2,w) = \left[\partial_w G_B(z_1,w) -
\partial_w G_B(z_2,w) \right]
\left[(\epsilon_1\epsilon_2)
\partial_{z_1}\partial_{z_2}G_B(z_1,z_2) + \right. \\
           &  & \quad + \sum_{j,i=1}^4 (\epsilon_1k_i)(\epsilon_2k_j)
\partial_{z_1}G_B(z_i,z_1) \partial_{z_2}G_B(z_j,z_2)] + \\
           & + & (\epsilon_1k_3) \sum_{j=1}^4(\epsilon_2k_j)
\partial_{z_2}G_B(z_j,z_2)I(z_1)
\left[\partial_w G_B(z_2,w) - \partial_w G_B(z_3,w)\right] + \\
           & - & (\epsilon_2k_3) \sum_{j=1}^4(\epsilon_1k_j)
\partial_{z_1}G_B(z_j,z_1)I(z_2) \left[\partial_w G_B(z_1,w) -
\partial_w G_B(z_3,w)\right] + \\
           & - & (\epsilon_1\epsilon_2)
\sum_{j=1}^4 \left[(k_2k_j) \partial_{z_2} G_B(z_j,z_2)\partial_w G_B(z_2,w)
B^+(z_1,w) + \right. \\
           &  & \quad - \left. (k_1k_j)
\partial_{z_1} G_B(z_j,z_1) \partial_w G_B(z_1,w)B^-(z_2,w) \right] + \\
           & - & \sum_{j,i=1}^4 (\epsilon_1k_i)(\epsilon_2k_j)
\left[\partial_{z_1} G_B(z_i,z_1)\partial_w G_B(z_j,w)B^+(z_2,w) + \right. \\
           &  & \quad - \left. \partial_{z_2} G_B(z_j,z_2)
\partial_w G_B(z_i,w)B^+(z_1,w) \right] + \\
           & + & \sum_{j=1}^4 \partial_w G_B(z_j,w)
\left\{ \left[(\epsilon_1k_2)(\epsilon_2k_j)-
(\epsilon_1\epsilon_2)(k_2k_j)\right]C_2(z_1,z_2,w) + \right. \\
           &  & \quad +\left. \left[(\epsilon_2k_1)(\epsilon_1k_j)-
(\epsilon_1\epsilon_2)(k_1k_j)\right] C_1(z_1,z_2,w) \right\} + \\
           & + & 2 \sum_{j=1}^4 \partial_w G_B(z_j,w)
\left[(\epsilon_1k_3)(\epsilon_2k_j)D_2(z_1,z_2,w) +
(\epsilon_2k_4)(\epsilon_1k_j) D_1(z_1,z_2,w)\right] + \\
           & + & \left[(\epsilon_1k_2)(\epsilon_2k_1) -
(\epsilon_1\epsilon_2)(k_1k_2)\right]
\left[\partial_w G_B(z_1,w) - \partial_w G_B(z_2,w)\right] \times  \\
           & & \quad \times \left[G^+(z_1,z_2)^2
+2G^+(z_1,z_2)G^-(z_1,z_2)\right] +\\
           & + & 2  \ G^+(z_1,z_2)G^-(z_1,z_2) \{(\epsilon_1k_2)
\sum_{j=1,3}(\epsilon_2k_j)\left[\partial_w G_B(z_2,w) -
\partial_w G_B(z_j,w)\right] +  \\
           &   & \quad+\left[(\epsilon_1\epsilon_2)(k_1k_2)+
(\epsilon_1k_3)(\epsilon_2k_1)\right] \left[\partial_w G_B(z_1,w) -
\partial_w G_B(z_3,w)\right]\},
\eean
\bean
{\cal A}_3(& z_1 &,z_2,w) = \partial_w G_B(z_1,w)[(\epsilon_2k_1)
\partial_{z_1} G_B(z_1,z_2)\sum_{j=1}^4(k_jk_2)\partial_{z_2}
G_B(z_2,z_j)+  \\
           &  & \quad - \sum_{j,i=1}^4 (k_1k_i)(\epsilon_2k_j)
\partial_{z_1}G_B(z_i,z_1)\partial_{z_2}G_B(z_j,z_2)] + \\
           & - & (k_1k_3)\sum_{j=1}^4 (\epsilon_2k_j)
\partial_{z_2} G_B(z_j,z_2)
\left[\partial_w G_B(z_2,w) - \partial_w G_B(z_3,w)\right]I(z_1)+\\
           & - & (\epsilon_2k_4)\partial_w G_B(z_1,w)
\sum_{j=1}^4(k_1k_j)\partial_{z_1} G_B(z_j,z_1)I(z_2)+ \\
           & - & \sum_{j=1}^4 \left[(k_1k_2)(\epsilon_2k_j)-
(\epsilon_2k_1)(k_2k_j)\right] \partial_w G_B(z_j,w)C_2(z_1,z_2,w)+ \\
           & + & B^+(z_1,w)\sum_{j=1}^4 \partial_{z_2} G_B(z_j,z_2)
 \left[(k_2k_j) (\epsilon_2k_1)\partial_w G_B(z_2,w) + \right. \\
           &  & \quad - \sum_{i=1}^4(\epsilon_2k_j)(k_1k_i)
\partial_w G_B(z_i,w)] + \\
           & + & 2 G^+(z_1,z_2)G^-(z_1,z_2) \left\{(\epsilon_2k_1)(k_1k_3)
\left[\partial_w G_B(z_3,w) - \partial_w G_B(z_1,w)\right] + \right. \\
           &  & \quad - \left. (\epsilon_2k_4)(k_1k_2)
\left[\partial_w G_B(z_3,w) - \partial_w G_B(z_2,w)\right] \right\}+ \\
           & - & 2 \sum_{j=1}^4 \partial_w G_B(z_j,w)
\left[(\epsilon_2k_4)(k_1k_j)D_1(z_1,z_2,w)+
(k_1k_3)(\epsilon_2k_j)D_2(z_1,z_2,w)\right],
\eean
\bean
{\cal A}_4(& z_1 &, z_2,w)= \partial_w G_B(z_2,w)[(\epsilon_1k_2)
\partial_{z_2}G_B(z_1,z_2)\sum_{j=1}^4(k_jk_1)\partial_{z_1}
G_B(z_1,z_j)+ \\
           &  & \quad - \sum_{j,i=1}^4(k_2k_i)(\epsilon_1k_j)
\partial_{z_1}G_B(z_j,z_1)\partial_{z_2}G_B(z_i,z_2)] + \\
           & - & (k_2k_3)\sum_{j=1}^4 (\epsilon_1k_j)
\partial_{z_1} G_B(z_j,z_1)\left[\partial_w G_B(z_1,w) -
\partial_w G_B(z_3,w)\right] I(z_2)+\\
           & + & (\epsilon_1k_4)\partial_w G_B(z_1,w)
\sum_{j=1}^4(k_1k_j)\partial_{z_1} G_B(z_j,z_1)I(z_2)+ \\
           & + & \sum_{j=1}^4 \left[(k_1k_2)(\epsilon_1k_j)-
(\epsilon_1k_2)(k_1k_j)\right] \partial_w G_B(z_j,w)C_1(z_1,z_2,w)+ \\
           & - & B^+(z_2,w)\sum_{j=1}^4 \partial_{z_1} G_B(z_j,z_1)
\left[\sum_{i=1}^4(\epsilon_1k_j)(k_2k_i)
\partial_w G_B(z_i,w) +\right. \\
           &   & \quad - (k_1k_j)(\epsilon_1k_2)
\partial_w G_B(z_1,w) \Big] \\
           & + & 2G^+(z_1,z_2)G^-(z_1,z_2)\left[(\epsilon_1k_3)(k_1k_2) -
(\epsilon_1k_2)(k_1k_3)\right]\times\\
           &   &\quad\times \left[\partial_w G_B(z_3,w) -
\partial_w G_B(z_2,w)\right] +\\
           & + & 2 \sum_{j=1}^4 \partial_w G_B(z_j,w)
\left[(\epsilon_1k_j)(k_1k_3) -
(k_1k_j)(\epsilon_1k_3)\right]D_1(z_1,z_2,w),
\eean
and finally
\bea
&B^{\pm}& \left[{}^{\alpha}_{\beta} \right] (z_i,w)=
\frac{{\cal I}(z_i)}{{\cal I}(w)}
\left[G^+(z_i,w) \pm  G^-(z_i,w)\right],\label{definition4}\\
&D_{1,2}& \left[{}^{\alpha}_{\beta} \right] (z_1,z_2,w) =
\frac{{\cal I}(z_{2,1})}{{\cal I}(w)}G^+(z_{1,2},w)G^-(z_1,z_2),\nonumber\\
&C_{1,2}& \left[{}^{\alpha}_{\beta} \right](z_1,z_2,w) =
\left[G^+(z_1,z_2)G^-(z_1,z_2) + \right. \nonumber\\
   & & \quad + \left. \frac{{\cal I}(z_{2,1})}{{\cal I}(w)}
G^+(z_{1,2},w) \left(G^+(z_1,z_2) + G^-(z_1,z_2)\right) \right] .
\eea

\begin{thebibliography}{999}
%
%
\bibitem{DP2} K. Aoki, E. D'Hoker and D.H. Phong, Nucl.Phys. {\bf B342}
              (1990) 149;\hfill\break
              E. D'Hoker and D.H. Phong, Phys.Rev.Lett. {\bf 70} (1993) 3692,
              Theor.Math.Phys. {\bf 98} (1994) 306 [hep-th/9404128],
              Nucl.Phys. {\bf B440} (1995) 24 [hep-th/9410152].
\bibitem{Ama} K. Amano, Nucl.Phys. {\bf B328} (1989) 510.
\bibitem{Mont} J.L. Montag and W.I. Weisberger, Nucl.Phys. {\bf B363}
              (1991) 527.
\bibitem{Gross} D. Gross, in "{\sl Proceedings of the Twenty-Fourth
                International Conference on High Energy Physics,
                Munich 1988 \/}, Springer Verlag, 1989.
\bibitem{Wein} S. Weinberg, ``{\sl Radiative corrections in string theory\/}'',
               Talk given at the APS meeting 1985, in DPF Conf. 1985, p. 850.
\bibitem{Fis} W. Fischler and L. Susskind, Phys.Lett. {\bf B171} (1986) 383,
              and {\bf B173} (1986) 262.
\bibitem{Tsey} A.A. Tseytlin, in ``{\sl Trieste Superstrings 1989\/}'',
               487, and references therein.
\bibitem{Bere} A. Berera, Nucl.Phys. {\bf B411} (1994) 157.
\bibitem{Kap} V.S. Kaplunovsky, Nucl.Phys. {\bf B307} (1988) 145
              [hep-th/9205070].
\bibitem{BK} Z. Bern and D. Kosower, Nucl.Phys. {\bf B379} (1992) 451.
\bibitem{DiV} P. Di Vecchia, A. Lerda, L. Magnea, R. Marotta and R. Russo,
              Nucl.Phys. {\bf B469} (1996) 235 [hep-th/9601143].
\bibitem{At} J.J. Atick, L.J. Dixon and A. Sen, Nucl.Phys. {\bf B292} (1987)
             109;\hfill\break
             J.J. Atick and A. Sen, Phys.Lett. {\bf 186B} (1987) 339,
             Nucl.Phys. {\bf B286} (1987) 189 and {\bf B293} (1987) 317.
\bibitem{PR0} A. Pasquinucci and K. Roland, Mod.Phys.Lett. {\bf A9}
             (1994) 3023  [hep-th/9405100].
\bibitem{PR1} A. Pasquinucci and K. Roland, Nucl.Phys. {\bf B440} (1995) 441
              [hep-th/9411015].
\bibitem{KLT} H. Kawai, D.C. Lewellen and S.H.H.Tye, Nucl.Phys.
              {\bf B288} (1987) 1.
\bibitem{Anto} I. Antoniadis, C. Bachas, C. Kounnas and P. Windey,
               Phys.Lett. {\bf 171B} (1986) 51; \\
               I. Antoniadis, C. Bachas and C. Kounnas,
               Nucl.Phys. {\bf B289} (1987) 87;\\
               I. Antoniadis and C. Bachas, Nucl.Phys. {\bf B298} (1988) 586.
\bibitem{Bluhm} R. Bluhm, L. Dolan and P. Goddard, Nucl.Phys. {\bf B309}
               (1988) 330.
\bibitem{KLT2} H. Kawai, D.C. Lewellen, J.A. Schwartz, S.H.H. Tye,
              Nucl.Phys. {\bf B299} (1988) 431.
\bibitem{FMS} D. Friedan, E. Martinec and S. Shenker, Nucl.Phys.
              {\bf B271} (1986) 93.
\bibitem{PR2} A. Pasquinucci and K. Roland, Nucl.Phys. {\bf B457} (1995) 27
              [hep-th/9508135].
\bibitem{PR3} A. Pasquinucci and K. Roland, Phys.Lett. {\bf B351} (1995) 131
              [hep-th/9503040].
\bibitem{Kaj} G. Cristofano, R. Marotta and K. Roland, Nucl.Phys.
              {\bf B392} (1993) 345.
\bibitem{DPh} For a review, see E. D'Hoker and D.H. Phong, Rev.Mod.Phys.
              {\bf 60} (1988) 917.
\bibitem{Kaj2} K. Roland, Phys.Lett. {\bf B312} (1993) 345.
\bibitem{Koste} V.A. Kostelecky, O. Lechtenfeld, W. Lerche, S. Samuel and
                S. Watamura, Nucl.Phys. {\bf B288} (1987) 173.
\bibitem{PR4} A. Pasquinucci and K. Roland, Nucl.Phys. {\bf B473} (1996)
              [hep-th/9602026]
\bibitem{PPlett} A. Pasquinucci and M. Petrini, hep-th/9708131.
\bibitem{Fay} J. Fay, ``{\sl Theta functions on Riemann Surfaces\/}'', Lecture
              Notes in Mathematics {\bf 352}, Springer-Verlag, Berlin, 1973;
              \hfill\break D. Mumford, ``{\sl Tata Lectures on Theta 1\/}'',
              Progress in Mathematics {\bf 28}, Birkh\"auser, Boston, 1983;
              \hfill\break
              J. Fay, Duke Math. J. {\bf 51} (1984) 109; \hfill\break
              J. Fay, Proceedings of Symposia in Pure Mathematics {\bf 49}
              (1989) 485.
\bibitem{Fey} L.M. Brown and R.P. Feynman, Phys.Rev. {\bf 85} (1952) 231.
\bibitem{PR5} A. Pasquinucci and K. Roland, Nucl.Phys. {\bf B485} (1997) 241
              [hep-th/9608022].
\bibitem{PRlon} A. Pasquinucci and K. Roland, in {\it Gauge Theories,
                Applied Supersymmetry and Quantum Gravity II}, proceedings
                of the workshop held at Imperial College, 5-10 July 1996,
                page 346, Imperial College Press, London, 1997
                [hep-th/9611013].
\bibitem{Mart} E. Martinec, Nucl.Phys. {\bf B281} (1987) 157.
\bibitem{PDV1} P. Di Vecchia, M.L. Frau, K. Hornfeck, A. Lerda, F.Pezzella
              and S. Sciuto, Nucl.Phys. {\bf B322} (1989) 317.
%
%


\end{thebibliography}
\end{document}